\documentclass{article} 
\usepackage{graphicx}  
\usepackage{epstopdf}

\usepackage{multicol}        
\usepackage{tikz}
\usepackage{paralist}
\usepackage{subfigure}

\usetikzlibrary{positioning}
\usetikzlibrary{arrows}
\usetikzlibrary{fit}

\newcommand{\kinectTM}{Kinect\texttrademark}
\newcommand{\kinectTMS}{\kinectTM~}

\newcommand{\sref}[1]{Sec.~\ref{#1}}
\newcommand{\fref}[1]{Fig.~\ref{#1}}

\newcommand{\eref}[1]{Eq.~\ref{#1}}

\newcommand{\aref}[1]{App.~\ref{#1}}
\renewcommand{\and}{\ifhmode\unskip\nobreak\fi\ $\cdot$\ }

\def\acknowledgement{\par\addvspace{17pt}\small\rmfamily
  \trivlist\if!Acknowledgements!\item[]\else
  \item[\hskip\labelsep\textsf{\bfseries Acknowledgements}]\fi%
}

\newcommand{\keywords}[1]{\par\addvspace\medskipamount{\rightskip=0pt plus1cm
  \noindent\textsf{\bfseries Keywords}\enspace\ignorespaces#1\par\addvspace\bigskipamount}}

\makeindex    
\begin{document}

\title{Eulerian vs. Lagrangian analyses of pedestrian dynamics asymmetries in a staircase landing}

\maketitle

\centerline{\scshape Alessandro Corbetta }
\smallskip
{\footnotesize
\centerline{Department of Applied Physics,}
\centerline{Eindhoven University of Technology,  The Netherlands}
\centerline{\texttt{a.corbetta@tue.nl}}
\medskip
} 

\medskip

\centerline{\scshape Chung-min Lee}
\smallskip
{\footnotesize
\centerline{Department of Mathematics and Statistics,}
\centerline{California  State University Long Beach, Long Beach, CA, USA}
\centerline{\texttt{chung-min.lee@csulb.edu}}
\medskip
}

\medskip

\centerline{\scshape Adrian Muntean}
\smallskip
{\footnotesize
\centerline{Department of Mathematics and Computer Science,}
\centerline{Karlstad University, Karlstad, Sweden}
\centerline{\texttt{adrian.muntean@kau.se}}
\medskip
}

\medskip

\centerline{\scshape Federico Toschi}
\smallskip
{\footnotesize
\centerline{Department of Applied Physics,}
\centerline{Department of Mathematics of Computer Science,}
\centerline{Eindhoven University of Technology, The Netherlands,}
  \centerline{CNR-IAC, Rome, Italy}
\centerline{\texttt{f.toschi@tue.nl}}

}

\bigskip

\begin{abstract}
Real-life, out-of-laboratory, measurements of pedestrian movements allow extensive and fully-resolved statistical analyses. However,  data acquisition in real-life is subjected to the wide heterogeneity that characterizes crowd flows over time. Disparate flow conditions, such as co-flows and counter-flows at low and at high pedestrian densities, typically follow randomly one another. 
When analysing the data in order to study the dynamics and behaviour of pedestrians it is crucial to be able 
  disentangle and to properly select (\textit{query})  data from statistically homogeneous flow conditions in order to avoid spurious statistics and to enable qualitative comparisons.

In this paper we extend our previous analysis on the asymmetric pedestrian dynamics on a staircase landing, where we collected a large statistical database of measurements from \textit{ad hoc} continuous recordings~\cite{corbettaTGF15}. This contribution has a two-fold aim: first, method-wise, we discuss two possible approaches to query experimental datasets for homogeneous flow conditions. For given flow conditions, we can either agglomerate measurements on a time-frame basis (Eulerian queries) or on a trajectory basis (Lagrangian queries). Second, we employ these two different perspectives to further explore asymmetries in the pedestrian dynamics in our measurement site. We report cross-comparisons of statistics of pedestrian positions, velocities and accelerations vs. flow conditions as well as vs. Eulerian or Lagrangian approach.
\end{abstract}

\medskip


\section{Introduction}        
\label{sect:intro}    
Experimental analyses of pedestrian dynamics for behavioural insights or predictive model validation saw a rapid proliferation over the last years~\cite{PhysRevLett.113.238701,johansson2010analysis,helbing2012crowd,johansson2007ACS,helbing2007dynamics}.
Fine-scale tracking-based data collections
have been growing in complexity and acquisition scales~\cite{zhang2014comparison,Brscic201477,mehner2015robust}, both in~\cite{DBLP:journals/ijon/BoltesS13,zhang2011transitions} and out~\cite{corbetta2014TRP,seer2014kinects,helbing2007dynamics} of controlled laboratory environments. 

Laboratory experiments enable detailed parametric studies of the crowd flow (see, e.g.,~\cite{schadschneider2009evacuation}), and may technically benefit from visual  markers to enhance automatic pedestrian detection and tracking~\cite{mehner2015robust}. Real-life condition measurements, recently tackled via, e.g.,  wireless sensors~\cite{roggen2011recognition} or, as here, via 3D sensors~\cite{corbetta2014TRP,seer2014kinects,DBLP:journals/ijon/BoltesS13}, likely eliminate potential behavioural biases by a laboratory environment (such as those introduced by the awareness of taking part into a scientific experiment).  However, they present hard automatic vision challenges~\cite{dalal2005histograms}.  In general, real-life measurements are realistically a must if one aims at resolved statistical descriptions of physical observables  (e.g., positions, velocities, accelerations) or quantification of related rare events~\cite{corbetta2016multiscale}. These descriptions are in fact possible by means of accumulating and agglomerating  data from continuous and long-time ranged measurements~\cite{corbetta2014TRP,Brscic201477}. Notably, extensive real-life measurements are  subjected to the natural uncontrollability and unpredictability of the crowd flow. In fact, in contrast with laboratory experiments, real-life measurements unavoidably include an  alternation of heterogeneous scenarios. A low density pedestrian flow can suddenly turn into a dense crowd, as it happens daily e.g. in a train station at rush hours~\cite{corbetta2016multiscale}. Likewise, scenarios where individuals walk undisturbed can alternate with group dynamics~\cite{zanlungo2014pedestrian}. Although data analyses including all traffic conditions at once are a possibility~\cite{Brscic201477} (e.g. to evaluate  global statistics or time-histories), inquiries on a homogeneous flow class basis, i.e. after a classification and selection of similar dynamics scenario, appear more useful toward a phenomenological understanding  of individual dynamics.  In other words, since we expect that pedestrians walking isolated from peers will exhibit a different dynamics than pedestrians walking in groups~\cite{PhysRevE.89.012811}, the agglomeration of data from these different scenarios appears to be`a logic step  for a (cross-)comparison analysis. Furthermore, modulo sufficiently long recording times, we can reach an arbitrary statistical resolution.
For such classification purposes, manual annotation has been often employed, e.g.  to select groups  in~\cite{PhysRevE.89.012811},  to classify walking patterns in~\cite{Tamura2013}, or to isolate people waiting in~\cite{seitzTGF15}. To the best of our knowledge, automatized agglomeration of homogeneous datasets from heterogeneous measurements ensembles is still an open problem, both technically and in terms of ``class homogeneity'' definition.  In the following, borrowing from the database terminology, we refer to selection operations as \textit{queries}.

In this paper we first discuss approaches for automatic selection of homogeneous flow data from heterogeneous long-term recordings. Then we apply these approaches to analyse and cross-compare massive pedestrian data collected by us in a year-long real-life measurement campaign at Eindhoven University of Technology, the Netherlands~\cite{corbetta2014TRP}. During this campaign we recorded on a  24/7 schedule  pedestrian trajectories  in a landing   (intermediate planar area between flights of stairs) with corridor-like geometry. We note that individuals walking in a landing are either ascending or descending the neighboring stair flights. This aspect, appearing on side of cultural preferences for the walking side~\cite{moussaid2009experimental}, induces asymmetries in the dynamics, which we discussed for selected flow conditions  in our previous work~\cite{corbettaTGF15}. Few experimental data have been collected in these scenarios, typically in the context of evacuation dynamics~\cite{hoskins2012differences,ronchi2014analysis,peacockmovement}. Natural heterogeneity in our data is high due to multiple natural traffic scenarios such as uni- or bi-directional flows with one or several pedestrians. From our analysis in~\cite{corbettaTGF15}, we expect that the number of pedestrians in the landing (taken as a surrogate of the density) and their walking directions strongly influence the dynamics. These two elements are at all insufficient in identifying a query. 
Processing extensive recordings querying for combinations of number of pedestrians and walking directions on a recording frame basis  appears to be a simple and  natural option. Nevertheless, long-range mutual interactions and memory effects are expected to influence the dynamics beyond single frames and rather steer entire trajectories. Queries selecting flow scenarios on a trajectory basis are hence a second, in a sense dual, standpoint.  Borrowing some known terminology from continuum mechanics, we define these queries respectively  ``Eulerian'' and ``Lagrangian''. Here we perform a cross-comparison of the statistics of pedestrian positions, velocities and accelerations in dependence on the different, but homogeneous, flow conditions. Furthermore, we employ selected  flow conditions to compare the two querying approaches.

This content of the paper is organized as follows: in \sref{sect:query} we formally introduce  the concepts of Eulerian and Lagrangian queries for pedestrian trajectories datasets. The data analyses of our dataset are the subject of \sref{sect:Meas}. The section includes a comparison of Eulerian and Lagrangian querying methodologies. A discussion  in \sref{sect:concl} closes the paper.

\section{Aggregation of homogeneous measurements: Eulerian and Lagrangian  queries}\label{sect:query}
\begin{figure}[t]
\center
\begin{tikzpicture}
\node[draw, circle,align=center] (P1) at (0,0) {P1\\$\rightarrow$};
\node[draw, circle,align=center, right=1cm of P1] (P2)  {P2\\$\leftarrow$};
\node[draw, circle,align=center, right=1cm of P2] (P3)  {P3\\$\rightarrow$};
\node[draw, circle,align=center, right=.5cm of P3] (P4)  {P4\\$\leftarrow$};
\draw[-] (P3)--(P4);
\node[below=.46cm of P1] (belP1) {};
\node[below=.46cm of P2] (belP2) {};
\node[below=.46cm of P3] (belP3) {};
\node[below=.46cm of P4] (belP4) {};
\draw[-,dashed] (P1)--(belP1);
\draw[-,dashed] (P2)--(belP2);
\draw[-,dashed] (P3)--(belP3);
\draw[-,dashed] (P4)--(belP4);
\node[below of=P1,left of=P1] (lstart) {};
\node[below of=P4,right of=P4] (lend) {$t$};
\draw[->,thick] (lstart)--(lend);
\node[draw, circle,align=center, below=1.5 cm of P1] (P5)  {P5\\$\rightarrow$};
\node[draw, circle,align=center, right=.75cm of P5] (P6) {P6\\$\leftarrow$};
\node[draw, circle,align=center, right=.5cm of P6] (P7)  {P7\\$\rightarrow$};
\node[below=1.85cm of P7] (belP7) {};
\draw[-,dashed] (P7)--(belP7);

\node[draw, circle,align=center, below right=.5cm of P6,fill=white ] (P8) {P8\\$\leftarrow$};
\draw[-] (P5)--(P6);
\draw[-] (P6)--(P7);
\draw[-] (P6)--(P8);
\draw[-] (P7)--(P8);
\node[below=5cm of P1,left of=P1] (lstart) {};
\node[below=5cm of P4,right of=P4] (lend) {$t$};
\draw[->,thick] (lstart)--(lend);
\node[below=1.85cm of P5] (belP5) {};
\node[below=1.85cm of P6] (belP6) {}; 

\node[below=.67cm of P8] (belP8) {};
\draw[-,dashed] (P5)--(belP5);
\draw[-,dashed] (P6)--(belP6);
\draw[-,dashed] (P8)--(belP8);

\end{tikzpicture}
\caption{We use graphs to represent Lagrangian selections of data. We associate each pedestrian  trajectory with a node in a graph carrying information on the direction. 
We connect with edges all those pedestrian/nodes that appear together in at least one time instant. The entrance time of each pedestrian define an order for the nodes.  P1 identifies a pedestrian going to the right that appears alone along the entire trajectory, i.e. undisturbed. P2 is undisturbed too, although going to the left. P3 and P4 have opposite direction, appear together at least in one time instant and do not appear with any other pedestrian. Cases P1, P2, P3, P4 are considered in \sref{sect:lagr}. A more complex scenario occurs for P5 -- P8. P5  enters first, before leaving he/she shares the landing for at least one time frame with P6, and,  afterwards, P6 appear together with both P7 and P8.\label{fig:graph-lagr-constr}}
\end{figure}
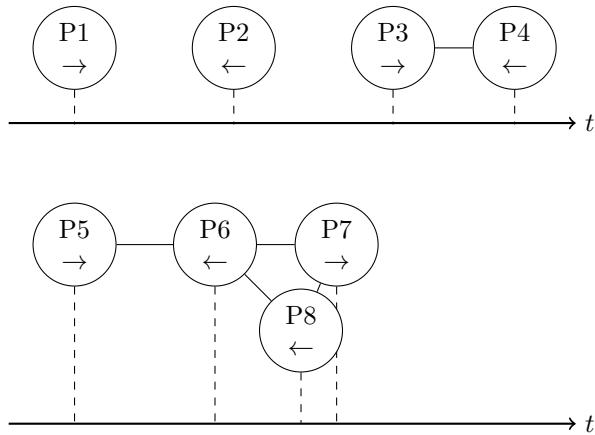

In this section we provide definitions and examples for Eulerian, i.e. frame-based, and Lagrangian, i.e. trajectory-based, data queries. 
Our definitions, although general, are here shaped after experimental scenarios like narrow corridors, as in our analyses in  \sref{fig:experiment-general}. In these cases, there are just two walking directions  and, consistently with the reference used in \fref{fig:experiment-general} these are from the left side to the right side or \textit{vice versa}. To be identified on a per scenario basis are the expected constituents of the dynamics:   the number of pedestrians involved  and the walking directions~\cite{corbettaTGF15}. In other words we expect a statistically similar (i.e. temporally homogeneous)  behaviour once fixed the number of pedestrians and given the walking direction. In order to better clarify the concept we provide here a few examples, anticipating the case studied for the following sections: 
\begin{enumerate}[(i)]
\item considering all the frames in which one pedestrian walks alone in our corridor in a given direction specifies an \textit{Eulerian} query;
\item generalizing (i), we can aggregate all time frames in which a given number of pedestrians with specified walking directions are in the facility. This is another \textit{Eulerian} query. 
\item case (i) includes frames with only one pedestrian. This implies that often only fragments of trajectories, possibly from heterogeneous flows, are included.  In fact, while entering the landing a pedestrian might initially be alone, but other pedestrians may appear successively. We label  as \textit{undisturbed} a pedestrian that is observed alone along the entire trajectory.  Isolating all the trajectories by undisturbed pedestrians (plus walking direction) implies a \textit{Lagrangian} query; 
\item the simplest avoidance scenario involves exactly two pedestrians, e.g. P3 and P4, (see \fref{fig:graph-lagr-constr}) walking in opposite direction. To ensure that the mutual presence is the only element influencing the two,  we require that no third pedestrian is present in the landing except for P3 and/or P4. Once more this is a \textit{Lagrangian} constraint as it pertains to the trajectories of P3 and P4 as a whole.
\end{enumerate}

\subsection{Lagrangian queries: interpretation and evaluation}\label{sect:lagrQ}

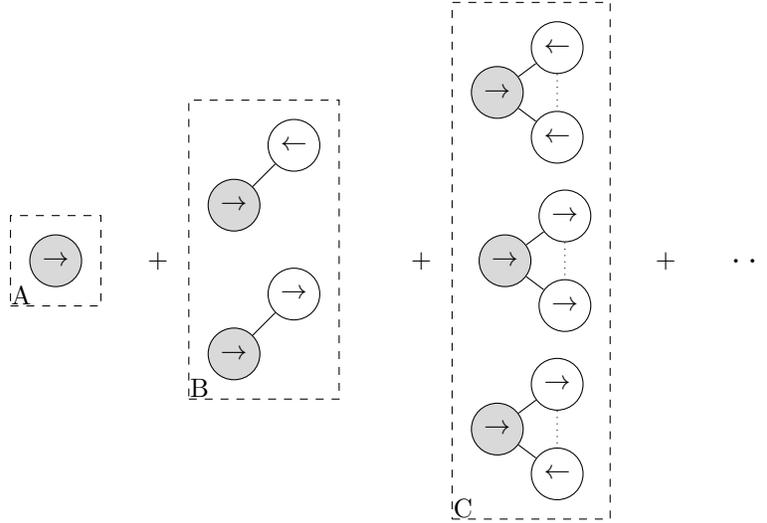
\begin{figure}[t!]
\center
\begin{tikzpicture}
\node[draw, circle,align=center,fill=gray!30] (P1) at (0,0) {$\rightarrow$};

\node (e1) [draw, dashed, fit= (P1), inner sep=0.25cm] {};
 \node [yshift=1.0ex, xshift=1.ex] at (e1.south west) {A};

\node[right=.75cm of P1] (P1pl) {$+$};

\node[draw, circle,align=center,above right=.25cm and .5 cm of P1pl,fill=gray!30] (P2) {$\rightarrow$};
\node[draw, circle,align=center,above right=0.3cm  and 0.3cm of P2] (P2o) {$\leftarrow$};
\draw[-] (P2)--(P2o);
\node[draw, circle,align=center,below right=.75cm and .5cm  of P1pl,fill=gray!30 ] (P3) {$\rightarrow$};
\node[draw, circle,align=center,above right=0.3cm  and 0.3cm of P3] (P3o) {$\rightarrow$};
\draw[-] (P3)--(P3o);
\node[right=4.25cm of P1] (P2pl) {$+$};

\node (e2) [draw, dashed, fit= (P2) (P2o) (P3) (P3o), inner sep=0.25cm] {};
 \node [yshift=1.0ex, xshift=1.ex] at (e2.south west) {B};

\node[draw, circle,align=center,below right=1.75cm and .5cm  of P2pl,fill=gray!30 ] (P4) {$\rightarrow$};
\node[draw, circle,align=center,above right=0.1cm  and 0.3cm of P4] (P4o) {$\rightarrow$};
\node[draw, circle,align=center,below right=0.1cm  and 0.3cm of P4] (P4o2) {$\leftarrow$};
\draw[-] (P4)--(P4o);
\draw[-] (P4)--(P4o2);
\draw[dotted] (P4o)--(P4o2);

\node[draw, circle,align=center,right= .5cm  of P2pl,fill=gray!30 ] (P5) {$\rightarrow$};
\node[draw, circle,align=center,above right=0.1cm  and 0.3cm of P5] (P5o) {$\rightarrow$};
\node[draw, circle,align=center,below right=0.1cm  and 0.3cm of P5] (P5o2) {$\rightarrow$};
\draw[-] (P5)--(P5o);
\draw[-] (P5)--(P5o2);
\draw[dotted] (P5o)--(P5o2);

\node[draw, circle,align=center,above right=1.75cm and .5cm  of P2pl,fill=gray!30 ] (P6) {$\rightarrow$};
\node[draw, circle,align=center,above right=0.1cm  and 0.3cm of P6] (P6o) {$\leftarrow$};
\node[draw, circle,align=center,below right=0.1cm  and 0.3cm of P6] (P6o2) {$\leftarrow$};
\draw[-] (P6)--(P6o);
\draw[-] (P6)--(P6o2);
\draw[dotted] (P6o)--(P6o2);

\node (e3)  [draw, dashed, fit= (P4) (P4o) (P4o2) (P5) (P5o) (P5o2)  (P6) (P6o) (P6o2) , inner sep=0.25cm] {};
 \node [yshift=1.0ex, xshift=1.ex] at (e3.south west) {C};

\node[right=7.5cm of P1] (P1pl) {$+$};

\node[right=8.5cm of P1] (P1pl) {\Large$\ldots$};

\end{tikzpicture}
\caption{Trajectories from  different Lagrangian scenarios contribute to the same Eulerian query. For instance, frames including just one pedestrian walking to the right, gathered in Eulerian sense, include contributions from heterogeneous sets of trajectories. Considering these trajectories (gray nodes) by the number of other pedestrians appeared, we have:
(A) trajectories from pedestrians walking alone to the right in Lagrangian sense (no further pedestrian appeared); (B) trajectories from  pedestrians that appear with a second pedestrian. The latter can have the same  or opposite (as P3-P4 in \fref{fig:graph-lagr-constr}) direction of the former; (C)  trajectories from  pedestrians appearing with two further pedestrians. These pedestrians may or may not appear together (thus an edge shall or shall not connect them, indicated with the dotted edge).   
\label{fig:graph-lagr-decomposition}}
\end{figure}

Lagrangian queries can be conveniently represented via  undirected graphs (we refer, e.g., to~\cite{bondy1976graph} for an introduction on graphs). We associate  each distinct pedestrian recorded to a graph node, including the pedestrian direction. Hence, we connect with an edge two nodes in case the two pedestrians appear together at least in one time frame. Analyzing the connected components of such a graph (i.e. those subgraph in which each node is connected to all others via a  path constituted of one or more edges~\cite{bondy1976graph}) we can extract homogeneous flow conditions with respect to trajectories. Pedestrians walking undisturbed are identified by all connected components with just one node  (e.g. P1 and P2 in \fref{fig:graph-lagr-constr} that have opposite directions). Scenarios involving just two pedestrians are identified by connected components with two nodes. Hence  avoidance  scenarios involving two pedestrians  (as in \sref{sect:lagr}) are defined by the connected components of the graph having two nodes associated to opposite directions  (cf. P3 -- P4 in \fref{fig:graph-lagr-constr}). Notably, this graph based selection comes at low computational costs as:
\begin{inparaenum}[(i)]
\item one pass of the dataset is sufficient to build the graph;
\item querying for connected components is a light operation on modern graph libraries such as~\cite{hagberg-2008-exploring}.
\end{inparaenum}
We can use this graph representation to interpret further the difference between Eulerian and Lagrangian queries. As an example, we consider the queries (i) and (iii) in \sref{sect:query}. Following (i) we isolate all the time frames in which one pedestrian walking in a given direction (e.g. from the left side of the corridor to the right side) is observed. Measurements from several trajectories fragments remain thus agglomerated.  In \fref{fig:graph-lagr-decomposition} we report a Lagrangian, graph based, classification of these trajectory fragments. 
There the number of nodes of the  connected components to which each trajectory fragment belongs gives the sorting criterion. Hence, the query (i) includes the entire selection given by (iii) (connected components with just one node) plus measurements from pedestrians that in previous or future time frames will appear with two, three or more other individuals. The following observations are due:
\begin{enumerate}[(A)]
\item Eulerian selections (e.g., (i)-(ii) in \sref{sect:query})  aggregate conditions having similar load and/or analogous usage patterns of the corridor; 
\item conversely, Lagrangian selections (e.g., (iii)-(iv) in \sref{sect:query}) identify specific physical scenarios focusing on the involved pedestrians  (cf. cases P1 -- P4 in \fref{fig:graph-lagr-constr}). In general these scenarios appear ideal references in social-force-like modeling perspectives~\cite{helbing1995PRE}, where the pedestrian motion is a sum of a \textit{desired} component in absence of other individuals (as in P1 or P2), plus additive term considering \textit{pair-wise} interactions (as for P3-P4). See \sref{sect:lagr} for further modeling considerations. 
\item The expansion in \fref{fig:graph-lagr-decomposition} grows with a super-exponential~\cite{stanley2001enumerative} number of different graph configurations as the number of considered pedestrians increases. This means that a graph based description may become impractical when many interacting pedestrians are considered and, for condition of high homogeneous crowding (co-flow of numerous pedestrians, counter-flows of two numerous groups), Eulerian queries may remain the only option. Nevertheless,  in these conditions, a prevalence of density-related effects over the Lagrangian graph edges seems reasonable. In other words, we expect strong similarities in the dynamics in case of large highly connected graphs, independently on the exact structure of the connections.
\end{enumerate}

\section{Asymmetric dynamics in a staircase landing}\label{sect:Meas}
We employ Eulerian and Lagrangian queries to select and analyze data from our large scale real-life measurements of pedestrian traffic in a corridor-shaped landing. For the sake of completeness we  report here a primer of the measurement campaign  and  we refer the interested reader  to~\cite{corbetta2014TRP,corbettaTGF15} for a more detailed overview of the traffic and to~\cite{corbetta2015MBE,corbetta2016multiscale} for the techniques employed.

\begin{figure}[t!h!]
  \begin{center}
     \scalebox{-1}[1]{ \includegraphics[width=.3\textwidth,trim=0.5cm .5cm 0cm 3.8cm,clip=true]{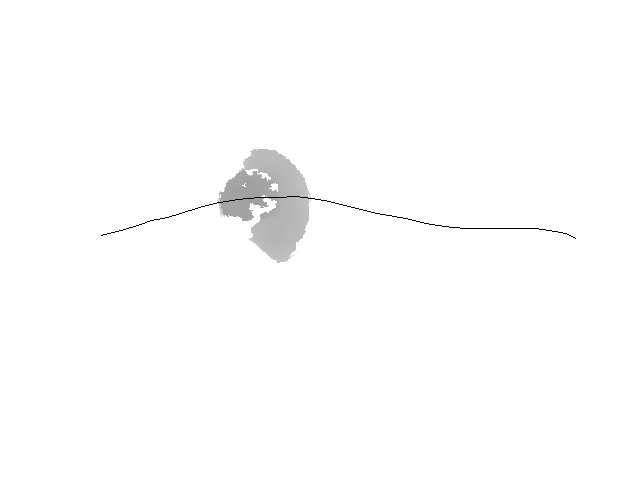} }
\includegraphics[width=.6\textwidth,trim=0 0cm 0cm .0cm,clip=true]{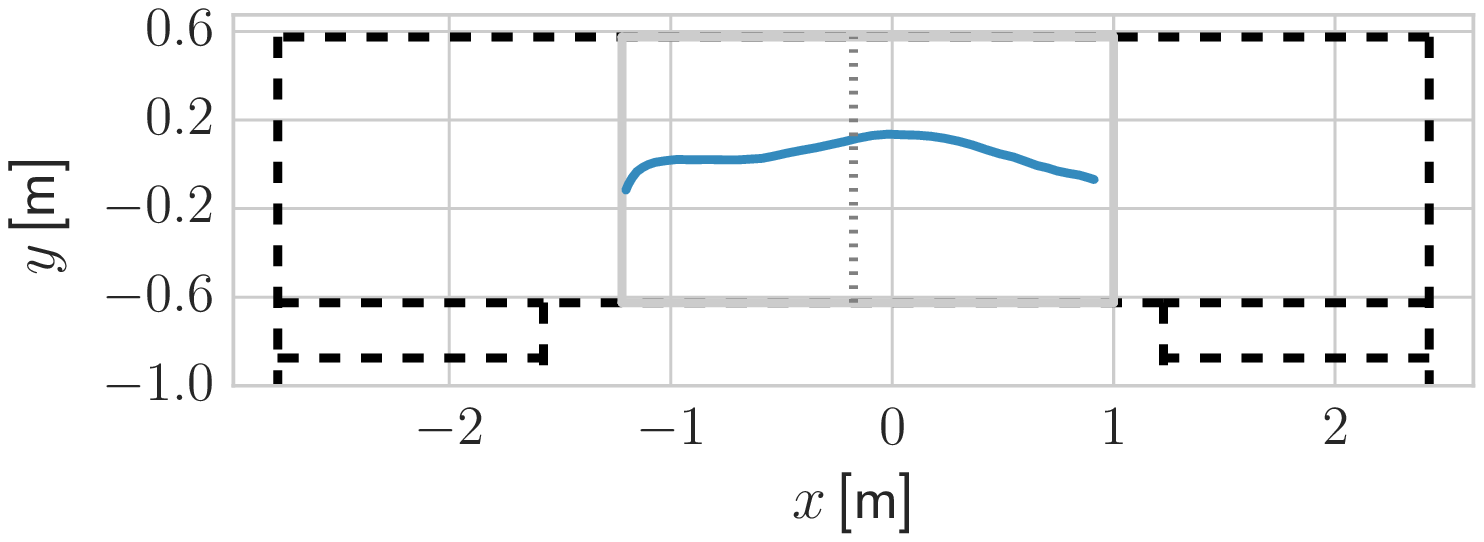}
 \scalebox{-1}[1]{ \includegraphics[width=.3\textwidth,trim=0.5cm .5cm 0cm 3.8cm,clip=true]{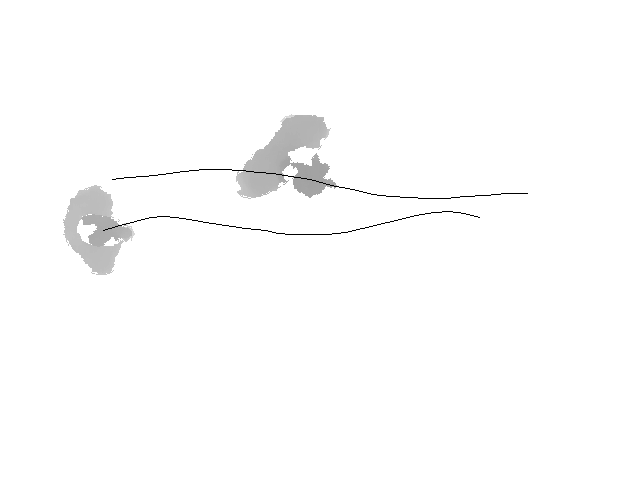} }
 \includegraphics[width=.6\textwidth,trim=0 0cm 0 0cm,clip=true]{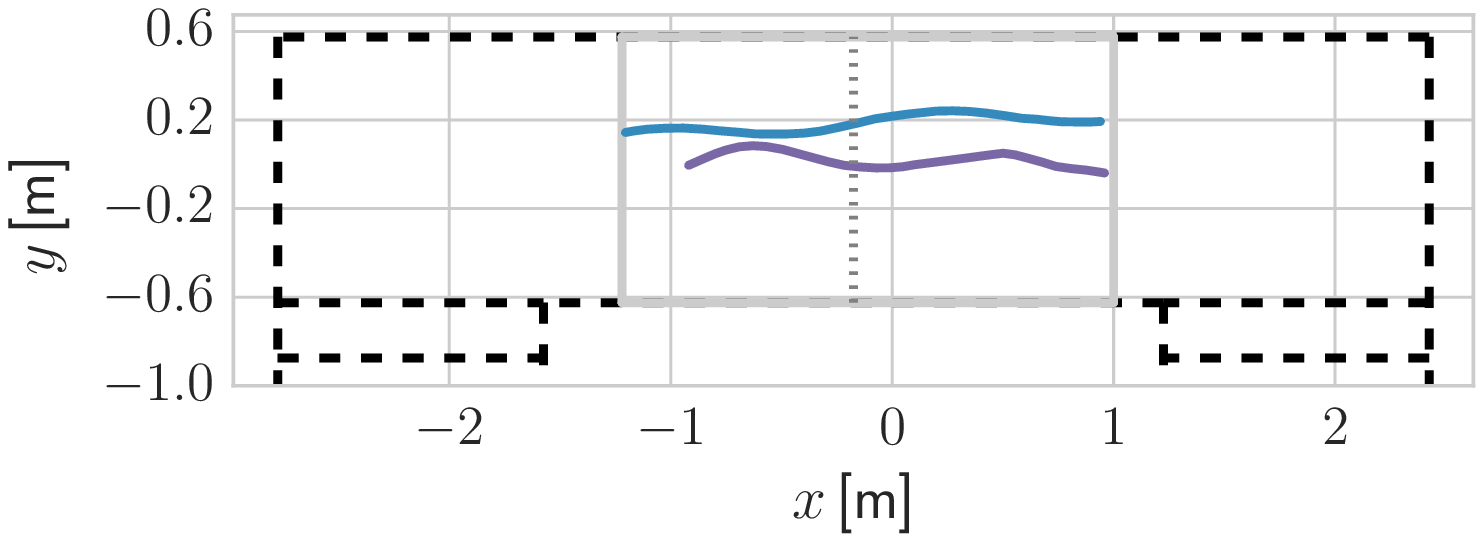}
 \scalebox{-1}[1]{ \includegraphics[width=.3\textwidth,trim=0.5cm .5cm 0cm 3.8cm,clip=true]{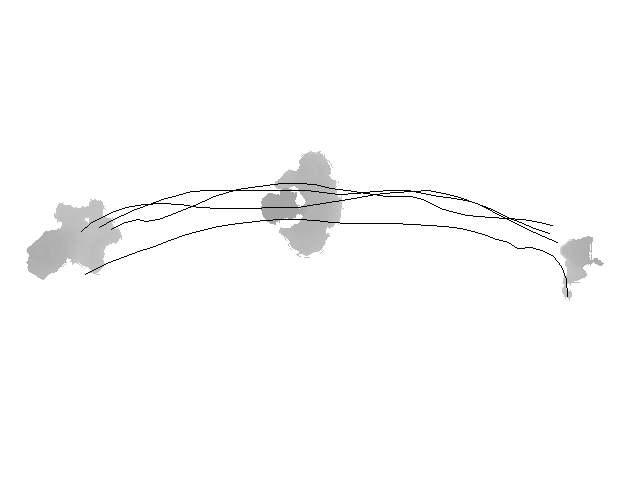} }
 \includegraphics[width=.6\textwidth,trim=0 0cm 0 0cm,clip=true]{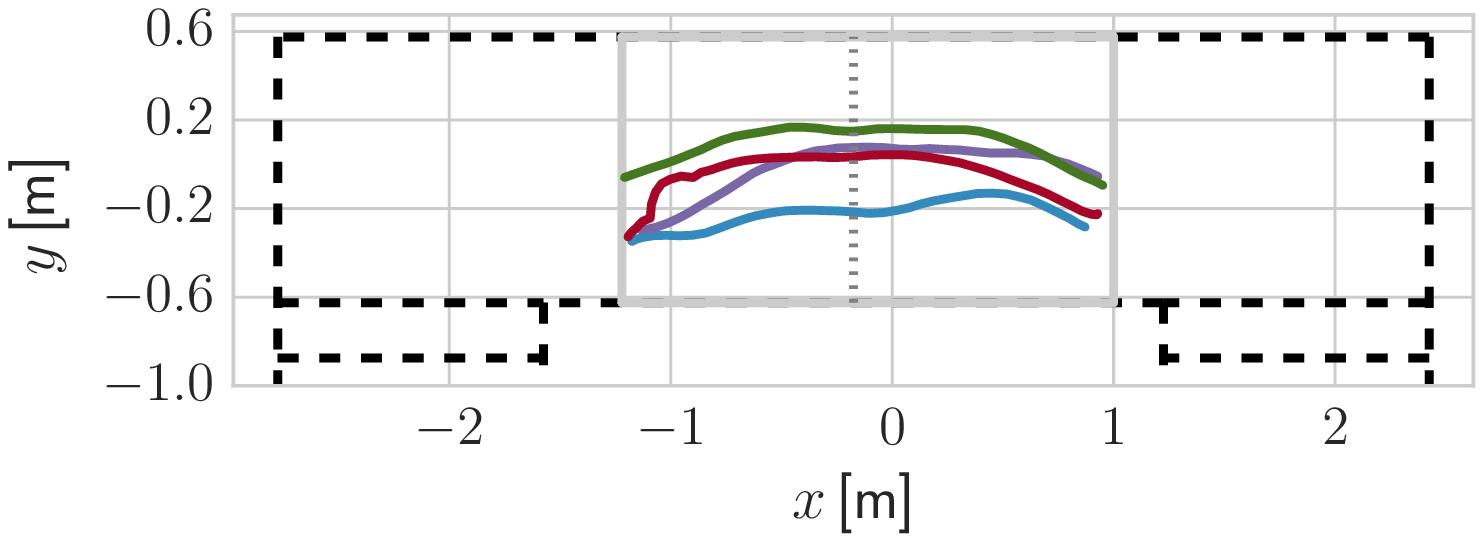}
\end{center}
\caption{  (Left panels) Depth frames taken in the landing by the \kinectTMS sensor, measured trajectories are superimposed. (Right panels) Measured trajectories in a sketch of the landing and considered $(x,y)$ reference. (Top row) One pedestrian walking from the left to the right hand side of the landing (2R) undisturbed. (Middle row) Two pedestrians moving from the right to left hand side of the corridor, i.e. co-flowing (2L). (Bottom row) A frame containing three pedestrians is plotted with all the four trajectories of the the connected component including the three of them (cf. \fref{fig:graph-lagr-constr} {P5 -- P8}).}
\label{fig:experiment-general} 
\end{figure}

In the one year starting from October 2013 we recorded via an overhead Microsoft \kinectTMS 3D-range sensor~\cite{Kinect} all pedestrians walking in  a landing within the Metaforum building at Eindhoven University of Technology. The landing connects two staircases in the configuration presented in the right panels of \fref{fig:experiment-general}, where individuals ascend in a clockwise direction from the ground floor to the first floor  of the building. The landing is $5.2\,$m long and $1.2\,$m wide, and the steps have the same width. Individuals at the ground floor reach the landing after $18$ steps, then they climb $4$ further steps arriving at the first floor.  Recordings went on a 24/7 basis and include data from 108 working days. With \textit{ad hoc} processing techniques of the \kinectTMS depth cloud and fluid mechanics-like tracking~\cite{OpenPTV,willneff2003spatio}, we collected \textit{ca.} 230,000 time-resolved high-resolution trajectories.

Trajectories span diverse flow scenarios, ranging from pedestrians walking undisturbed to clogged counter-flows. In the next subsections we analyze statistics from these flow scenarios employing Eulerian and Lagrangian queries.
In this section  we pursue an analysis of the dynamics as well as a comparison of the Eulerian and Lagrangian querying approaches.

\subsection{Eulerian overview of the dynamics}\label{sect:eulerian}

The U-shape of the landing influences the dynamics of pedestrians that follow curved trajectories to reach the staircase at the opposite end of the walkway   (cf. trajectories samples in \fref{fig:experiment-general}). Considering the stairs flights, pedestrians are furthermore ``globally'' ascending or descending through the building. For convenience we indicate the walking direction that allows one to ascend to the first floor as \textit{left to right} (2R, for brevity) and as \textit{right to left} (2L) the opposite case. Shape plus ``functional'' differences among walking direction allow the emergence of asymmetries in the dynamics for and within the different \textit{flow conditions}  (undisturbed pedestrian vs. multiple pedestrians vs. direction. Cf. also our previous work~\cite{corbettaTGF15}).

\begin{figure}[h!]
\includegraphics[width=.95\textwidth]{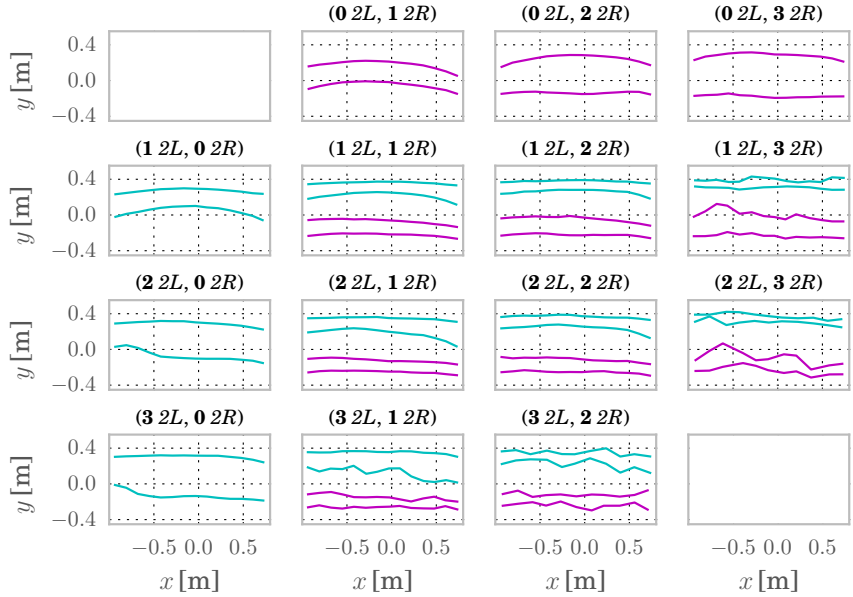}
\caption{Bands indicating the preferred pedestrian positions  in different flow conditions. Each plot reports different Eulerian queries, in the database of our measurements, based on the number of pedestrians traversing the corridor from right to left (2L) and from left to right (2R) and their ultimate direction.  Hence, in the subplot $(\textbf{N}\ \textit{2L}, \textbf{M}\ \textit{2R})$, we isolated all the frames containing $N$ pedestrians going to the left and $M$ going to the right. The cyan and the magenta lines limit, respectively, the preferred position bands  for pedestrians going to the left and to the right.  It is observed that the pedestrians conforms to the driving side preference by walking on the relative right side of the corridor, at least in the cases of counter-flowing.  Increase of co-flowing pedestrians results in expansion of the preferred position band in the transversal direction, but increase of counter-flowing pedestrians constricts the width of the preferred position band. }
 \label{fig:pref-path-combo}
\end{figure}

First we give an overview of the dynamics spanning over observed flow conditions adopting the Eulerian standpoint (cf. (A) and (C) in \sref{sect:lagrQ}).
Curved pedestrian trajectories fall preferentially in narrow curved bands that we use to compare our queries. The quantitative definition of the bands rely on binning the pedestrian position data according to the span-wise ($x$) position and taking statistics on the transversal position $y$. The bands reported in \fref{fig:pref-path-combo} range from the $15^{\rm th}$ to $85^{\rm th}$ percentiles of the pedestrians transversal positions  (cf. \aref{app:tech} for technical details).

Pedestrians maintain a relative right position in the corridor and in all flow configurations.   As the number of co-flow pedestrians increases, the width of the preferred position band increases, and, as the number of counter-flow pedestrians increases, the preferred position band becomes narrower.  As intuition suggests, the widths of preferred position bands in the counter-flow situation roughly follows the ratio between the numbers of pedestrians in both directions.  Note that for the 5 pedestrians cases when a group of 3 pedestrians and a group of two pedestrians going toward each other ((3 2L, 2 2R) and (2 2L, 3 2R) cases in \fref{fig:pref-path-combo}) the statistics are low so the boundaries of the preferred position bands are less smooth.

For an overview of the average walking velocities across these Eulerian queries we refer the reader to Sect. 3 in~\cite{corbettaTGF15}.

\subsection{Eulerian vs. Lagrangian queries of diluted flows}\label{sect:eu_vs_lagr}

Via the  queries in \sref{sect:eulerian} we  agglomerated our measurements following the Eulerian standpoint, and we showed position preferences and dynamics asymmetries. As commented in \sref{sect:lagrQ}, Eulerian queries exchange querying simplicity for physical clarity. When we consider trafficked dynamics involving many pedestrians, because of the combinatorial explosion of the Lagrangian graph configurations, Eulerian queries are likely to be the only option. However they mix heterogeneous physical scenarios.  In this section we compare results from Eulerian and Lagrangian queries when selecting flow conditions involving few pedestrians (one or two), i.e. ``diluted'' flows in our landing. 
Our analysis compares both bands of preferred position and speed fields.

\begin{figure}[thp!] 
\begin{center}
\subfigure[]{\includegraphics[width=.48\textwidth]{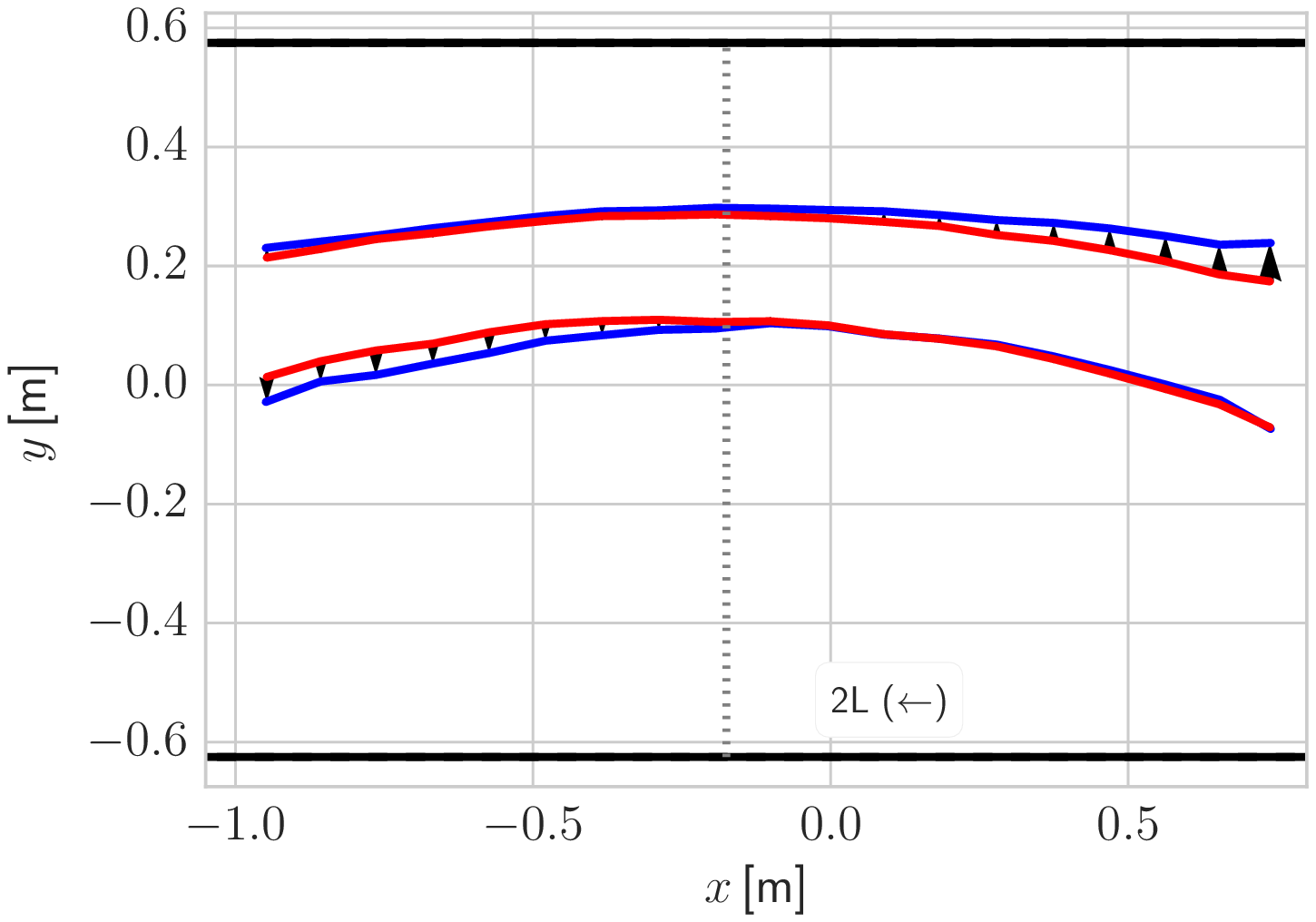}} 
\subfigure[]{\includegraphics[width=.48\textwidth]{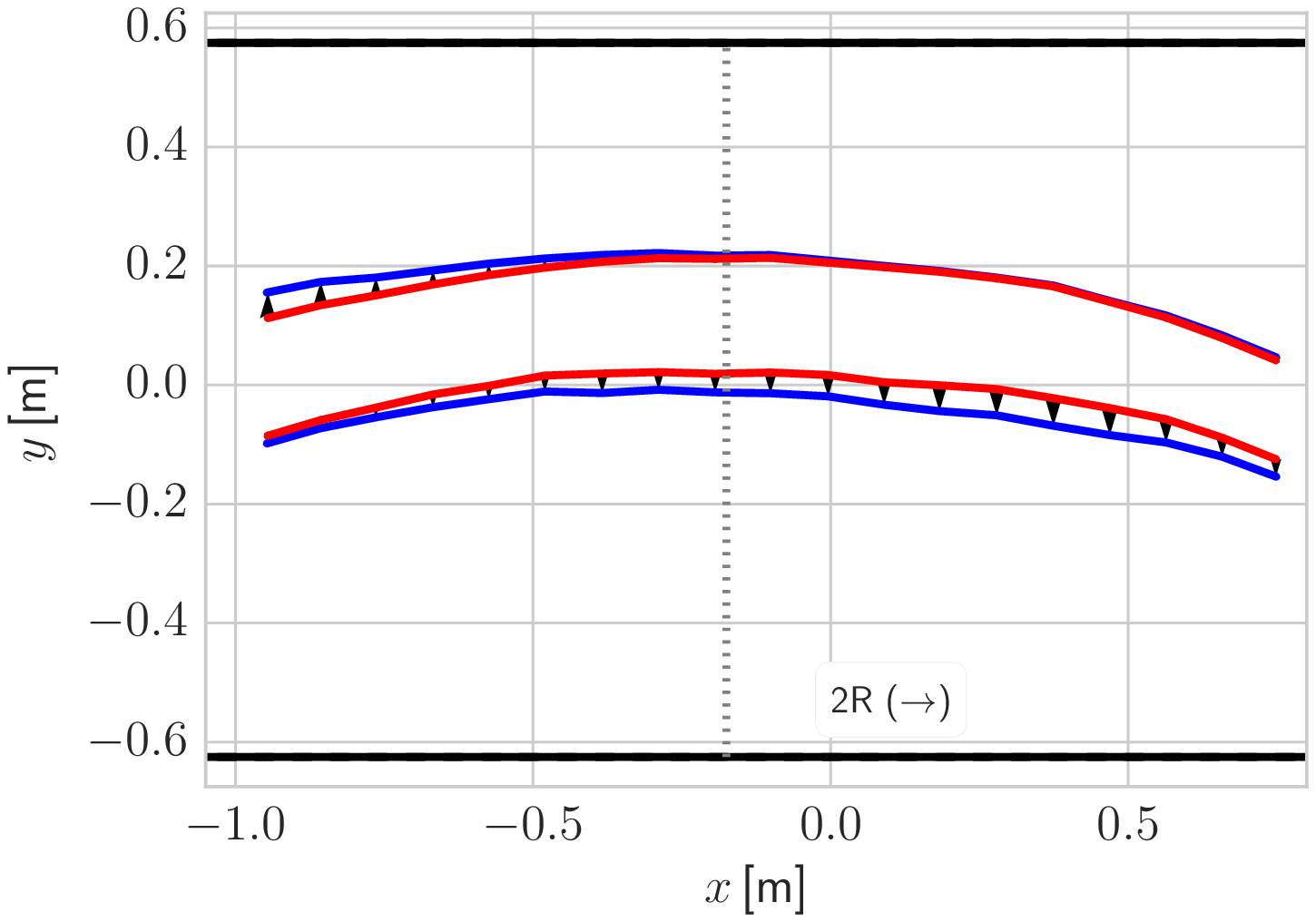}} 
\end{center}
\caption{Preferred position bands for pedestrians appearing alone in a frame.  The definition of these layers is discussed in \aref{app:tech}.   Preferred position bands for the 2L (panel a) and 2R (panel b) cases.  Red lines indicate the preferred position bands calculated from Lagrangian queries, and blue lines come from Eulerian queries.  The black arrows indicate the change from the Lagrangian queries.   Black vertical dotted lines indicate longitudinal center of the landing. The preferred position bands from Lagrangian queries are roughly symmetric with respect to the  longitudinal center, but in Eulerian queries such symmetry is lost.  Notably, the results from Eulerian queries have a larger width.}
\label{fig:singlePedEuLa}
\end{figure}

In \fref{fig:singlePedEuLa} we report the bands of preferred positions according to Eulerian and Lagrangian queries of single pedestrians for the two possible pedestrian directions (cf. cases (i) and (iii) in \sref{sect:query}). Consistently with \fref{fig:pref-path-combo} (subplots (1 2L, 0 2R) and (0 2L, 1 2R)), the 2L and 2R preferred position bands appear as being vertical translations by about 20 cm of each other with the 2L pedestrians walking on the upper side of the corridor, conforming to the driving side preference.  Although the relative position of the layers conforms with the cultural habit of keeping the driving side (cf. e.g.~\cite{moussaid2009experimental}), an influence of the landing geometry cannot be excluded. In fact the shape of the landing limits the sight on the staircases, hence right-hand side positions  may be kept to ease potential collisions (cf. \fref{fig:2counterflow} and  \fref{fig:interaction-force}). We note that bands from Lagrangian queries are symmetric to the longitudinal  center of the corridor, while this does not happen in the Eulerian case.  Specifically,  the entrance end of the bands expands more to the upper side of the corridor, and the exit end expands into the lower side of the corridor.

\begin{figure}[htp!]
\begin{center}
\subfigure[]{\includegraphics[width=.9\textwidth,trim=0 2.2cm 0 0,clip=true]{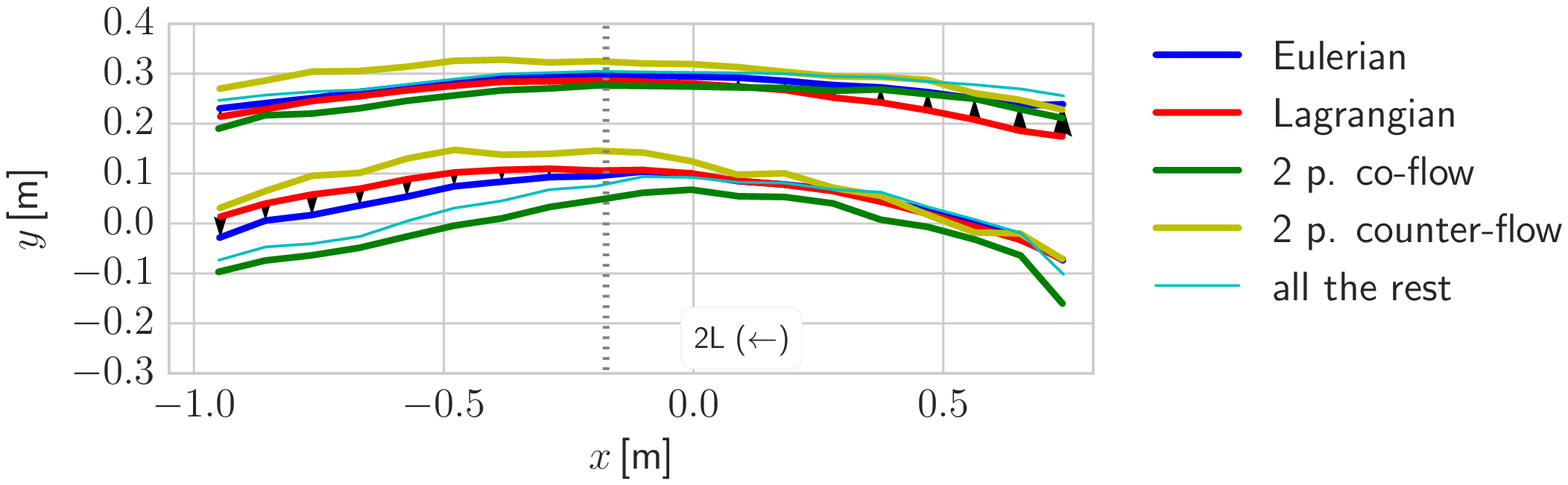}} 
\subfigure[]{\includegraphics[width=.9\textwidth,trim=0 2.2cm 0 0,clip=true]{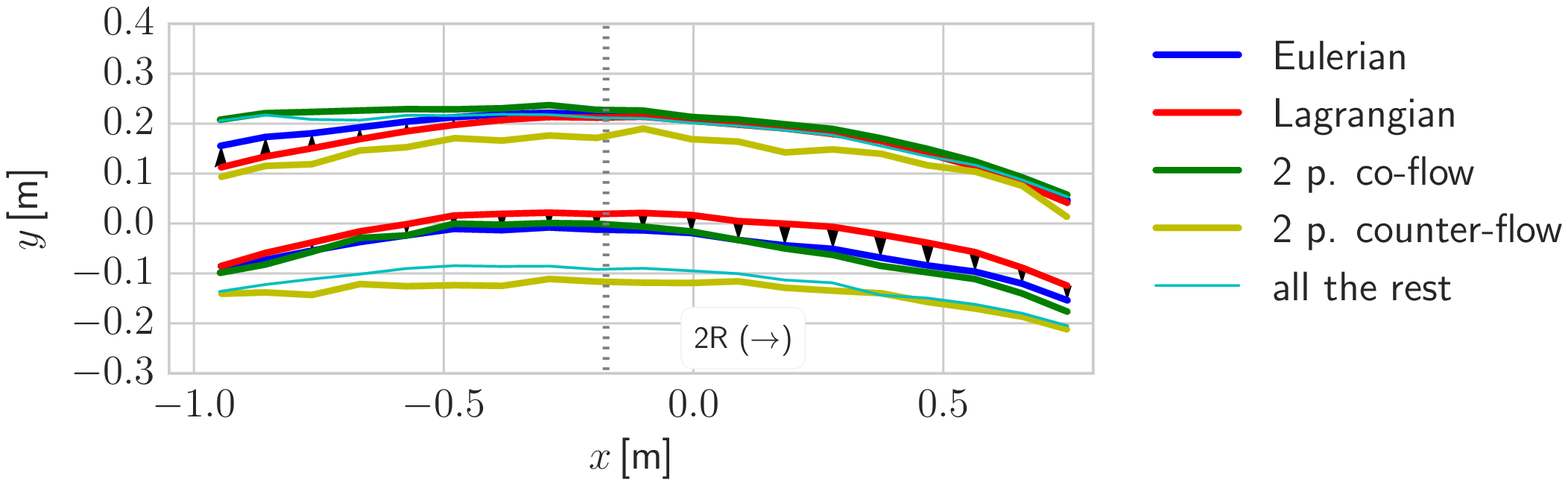}} 
\subfigure[]{\includegraphics[width=.49\textwidth]{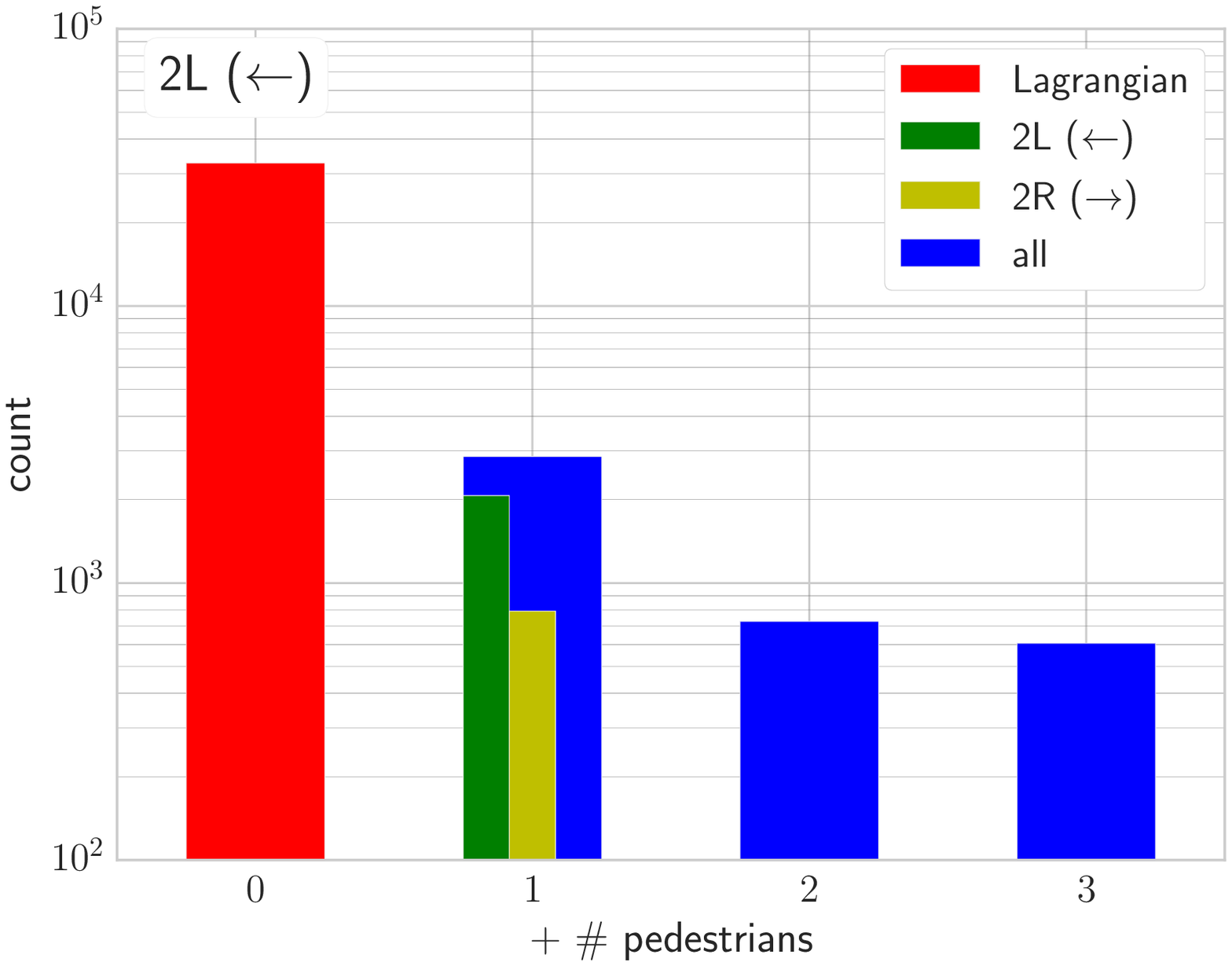}} 
\subfigure[]{\includegraphics[width=.49\textwidth]{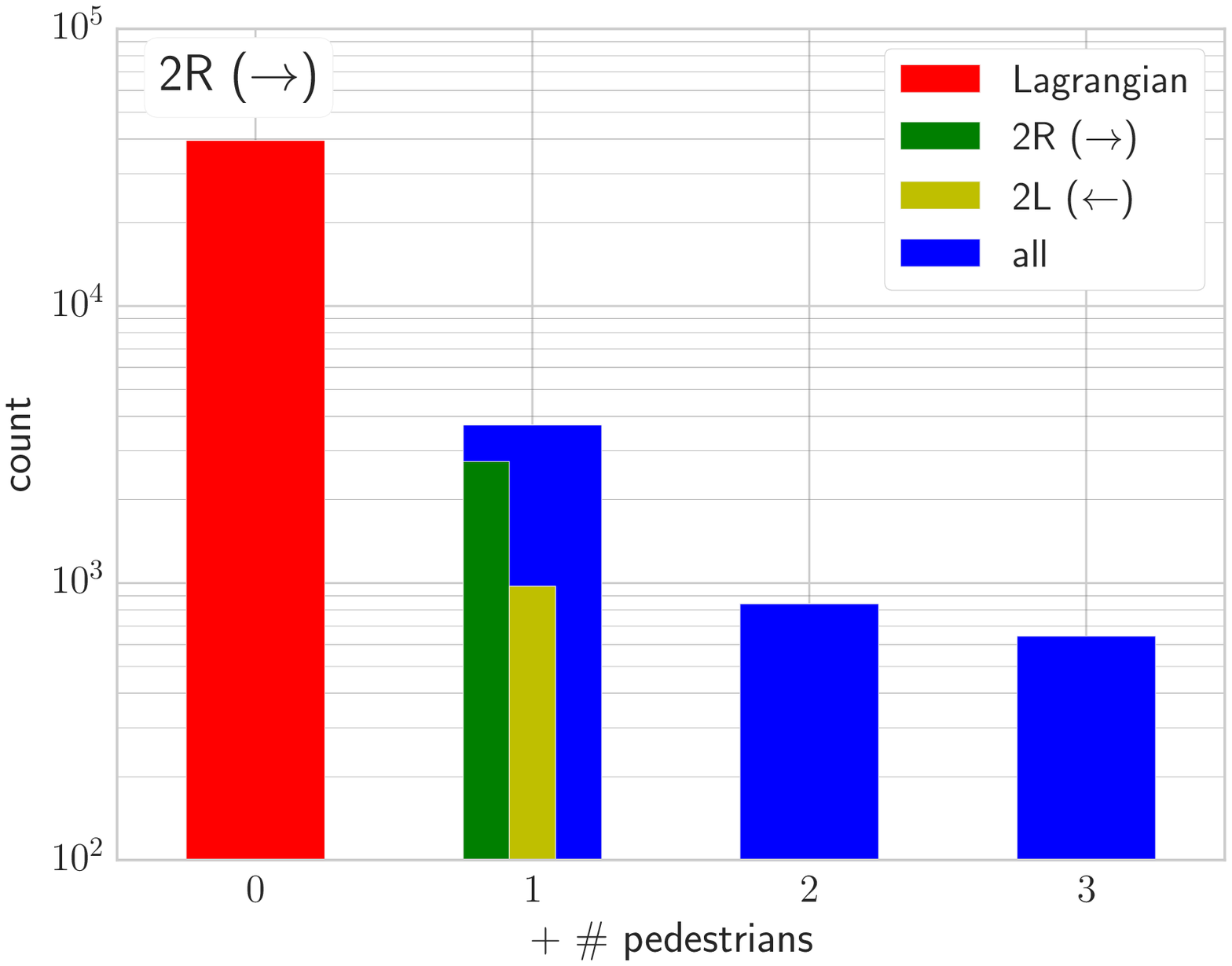}}  
\end{center}
\caption{ Contributions from different flow configurations to the preferred position bands in \fref{fig:singlePedEuLa}.  Panels (a) and (b) indicate, respectively, the preferred position bands from Eulerian queries (blue lines), Lagrangian queries of undisturbed pedestrians (red lines), two pedestrians co-flow (green lines), two pedestrians counter-flow (yellow lines) and the rest (cyan lines).  The co-flowing pedestrians (green lines) occupy a wider position band than other flow conditions while counter-flowing pedestrians (yellow lines) have preferred position bands shifted toward the relative right-hand side.  Panels (c) and (d) corresponds to 2L and 2R cases, respectively, the number of pedestrians in each flow configurations included in the Eulerian  queries.  Color scheme is the same as that in panels (a, b).  The largest contribution to the results of Eulerian queries other than the undisturbed pedestrians comes from co-flowing pedestrian pairs.  Hence at the entrance side  (right end for 2L, left end for 2R) the preferred position band is wider in Eulerian queries than in Lagrangian queries.}
\label{fig:singlePedEuLaPath-decomp}
\end{figure} 

The difference between the preferred position bands from the two queries comes from pedestrians who have met and will meet other pedestrians during their walk in the corridor (although currently alone).  \fref{fig:singlePedEuLaPath-decomp}{a} and \fref{fig:singlePedEuLaPath-decomp}{b} shows that the pair co-flow pedestrians occupy a wider preferred position band, and they are the largest group of single pedestrians eventually or previously sharing the corridor with others (cf. \fref{fig:singlePedEuLaPath-decomp}{c} and \fref{fig:singlePedEuLaPath-decomp}{d}).  For the increased width of the position bands at the exit end (right end for 2R, left end for 2L) in the Eulerian queries, it is possibly due to a combination of co-flow and counter-flow pedestrians following and avoiding others.

\begin{figure}[ht!]   
\begin{center}
\subfigure[]{\includegraphics[width=.4825\textwidth,trim=0 .2cm 2.6cm 0, clip=true]{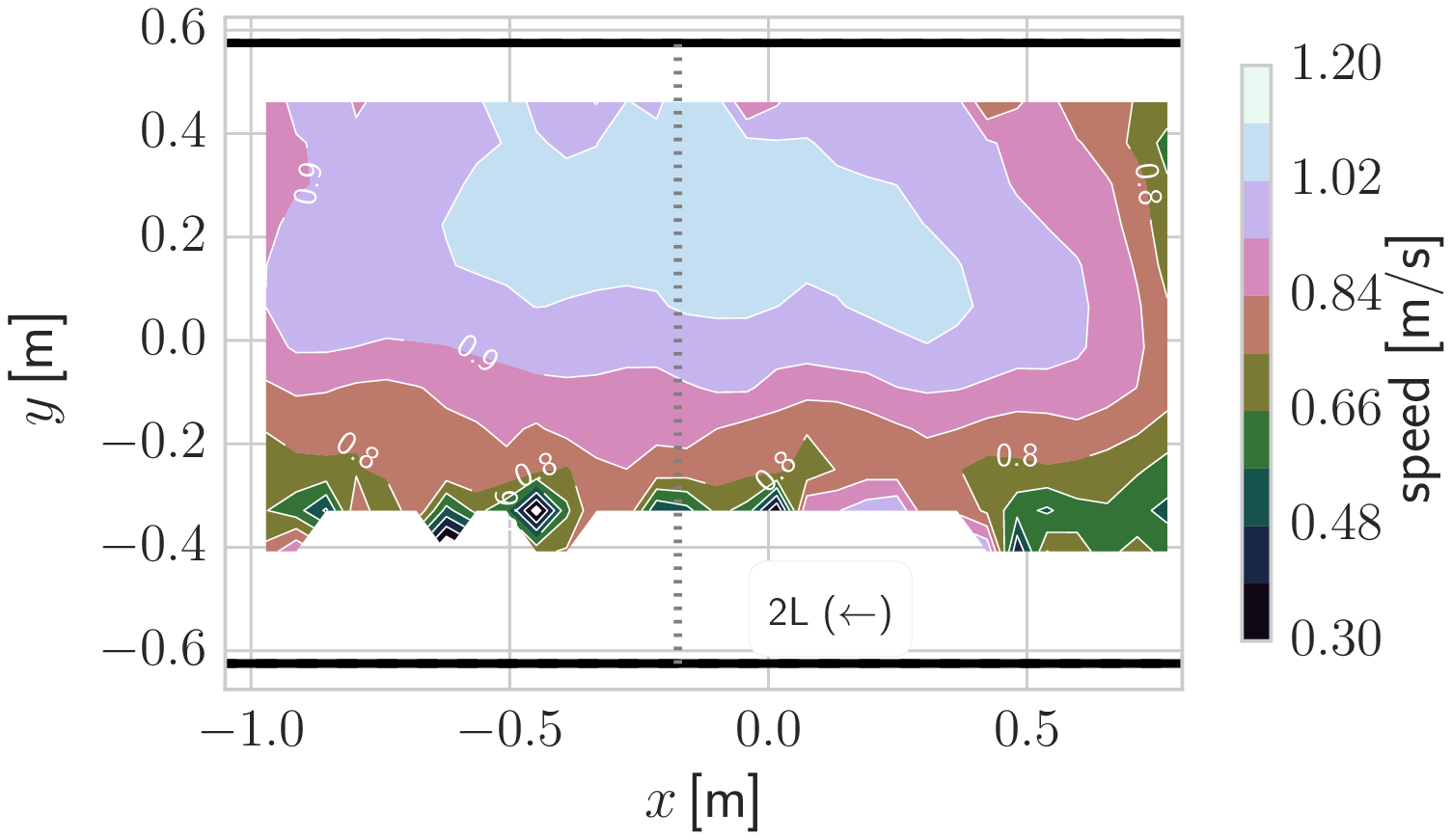}}
\subfigure[]{\includegraphics[width=.49\textwidth, trim=2.4cm .2cm 0cm 0, clip=true]{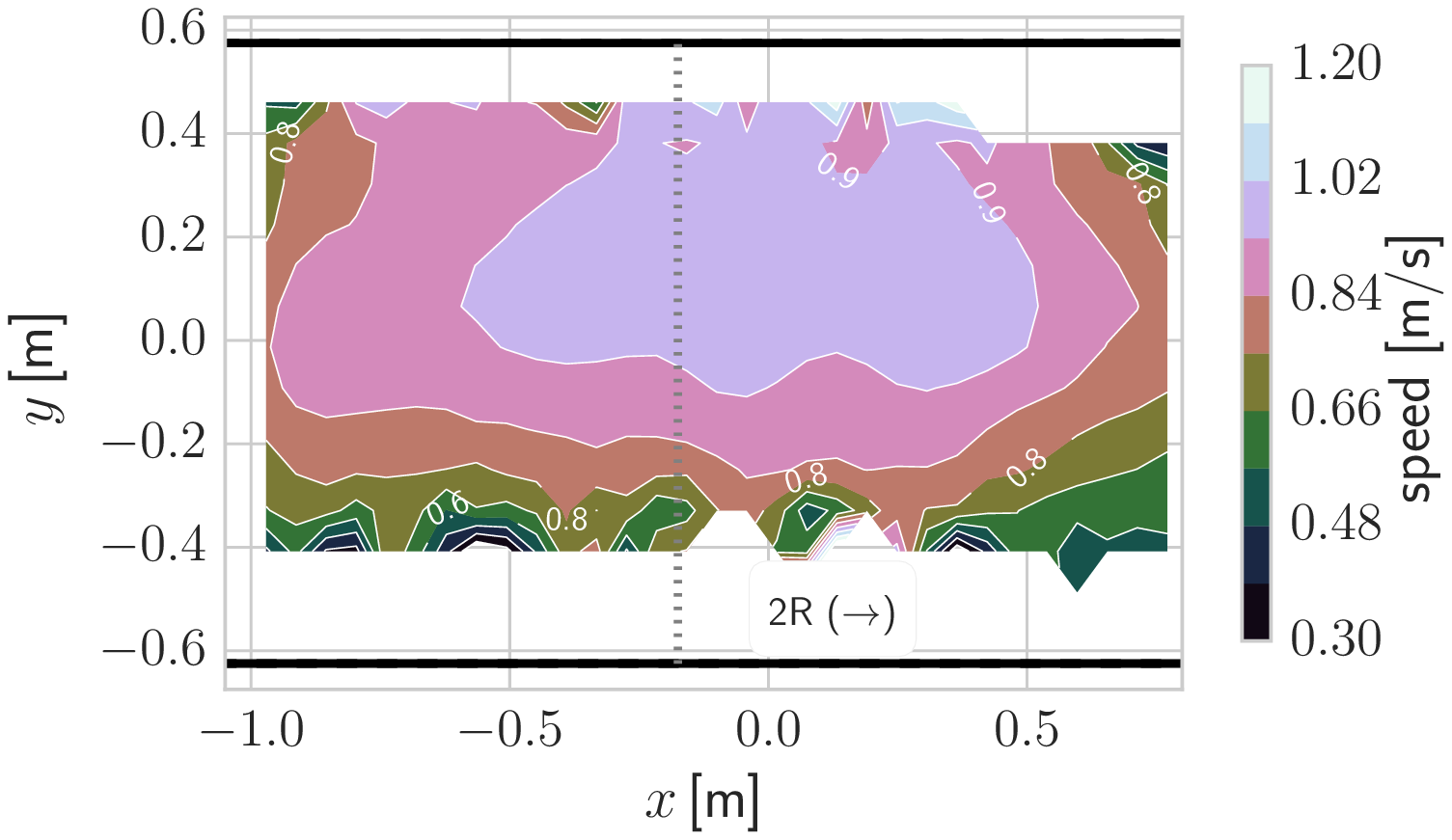}}
\subfigure[]{\includegraphics[width=.4825\textwidth,trim=0 .2cm 2.6cm 0, clip=true]{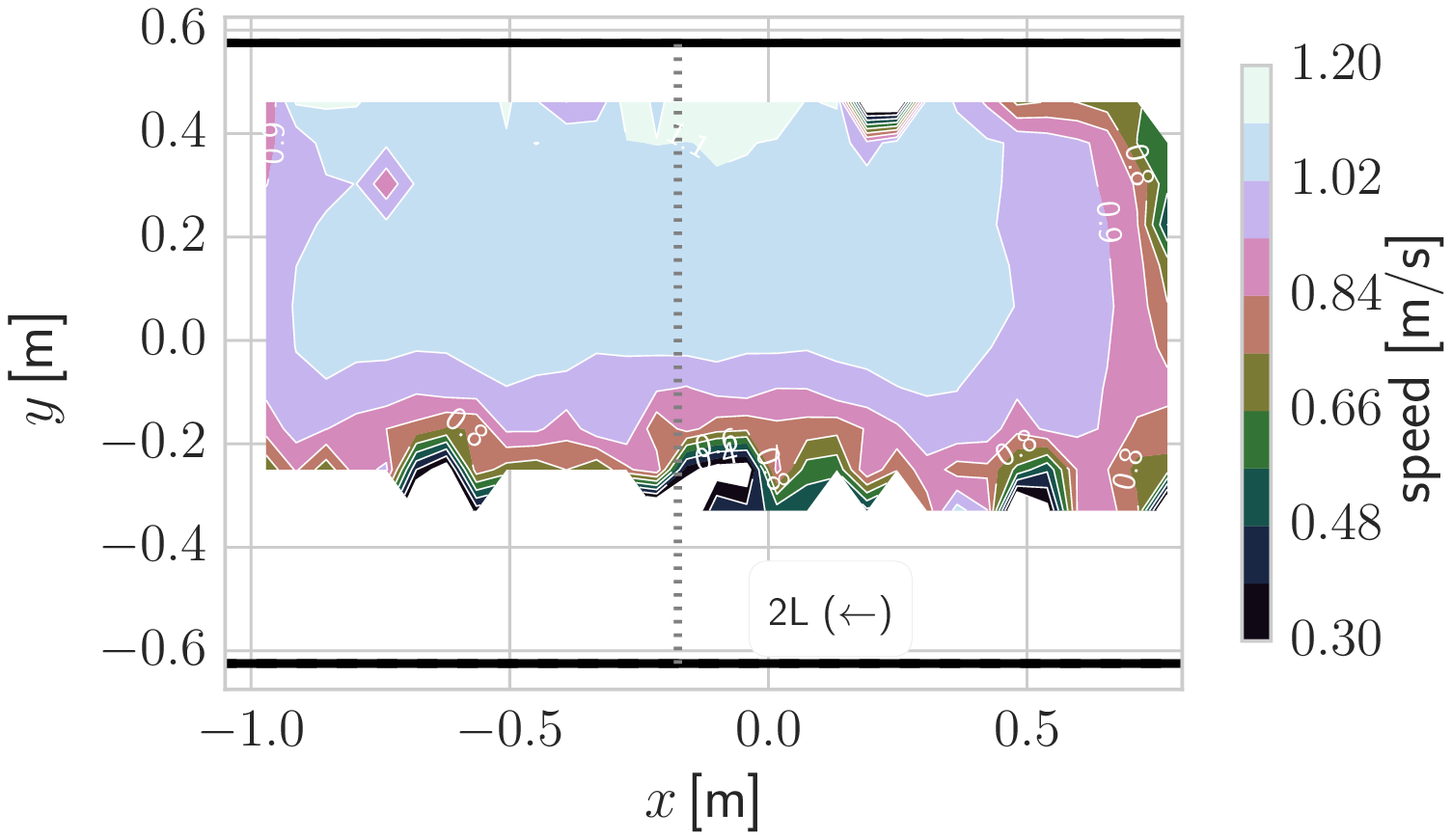}}
\subfigure[]{\includegraphics[width=.49\textwidth, trim=2.4cm .2cm 0cm 0, clip=true]{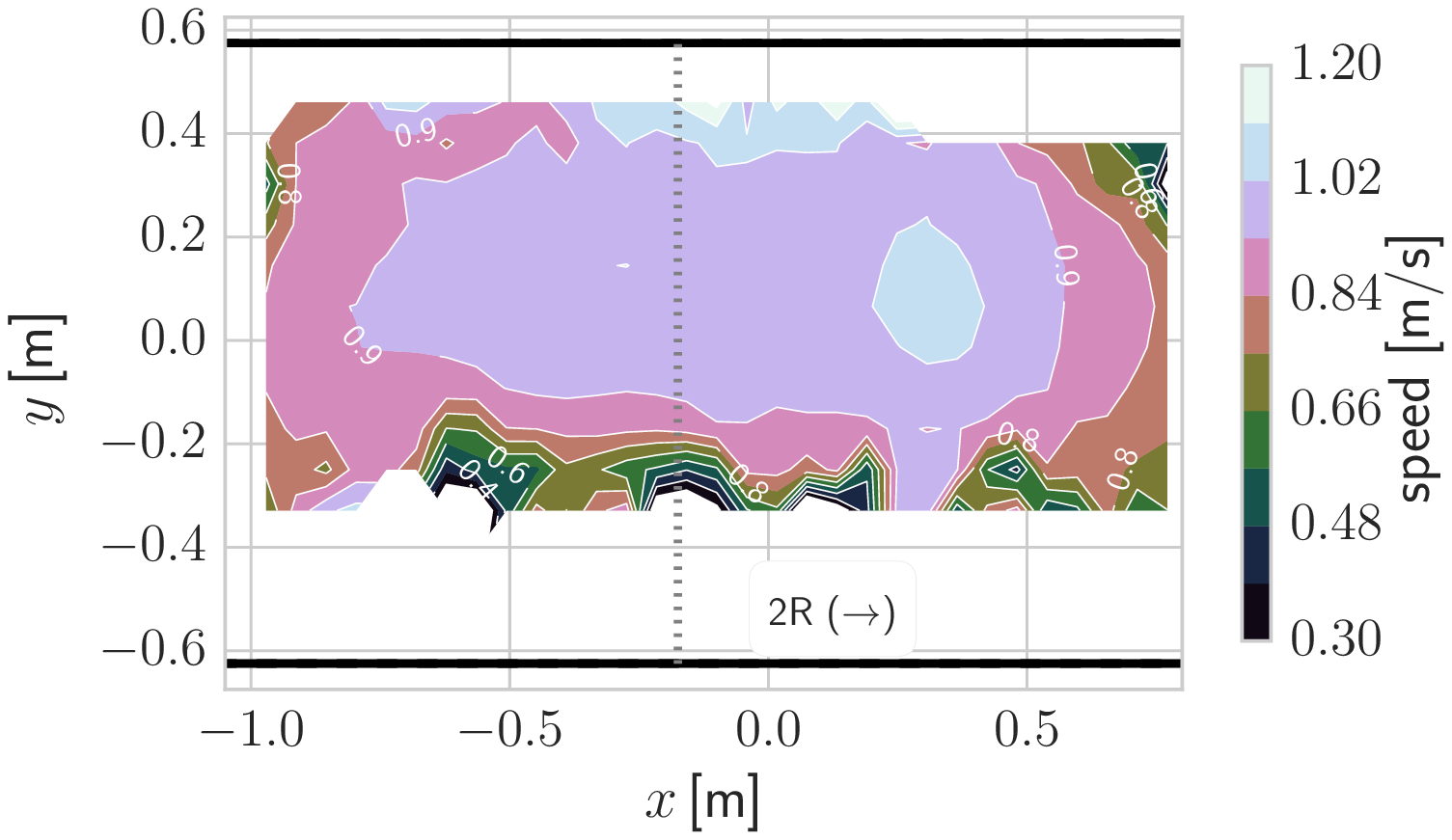}}
\end{center}
\caption{Spatial fields of pedestrian average walking speed considering pedestrians walking alone in Eulerian and Lagrangian sense. Panels (a, b) show average velocity fields for 2L and 2R pedestrians, respectively,  from Eulerian queries.  Panels (c, d) show average velocity fields for 2L and 2R pedestrians from Lagrangian queries.  In both 2L and 2R cases, pedestrian average speeds are lower in the Eulerian queries.  
}
\label{fig:speedEuLa} 
\end{figure}

\fref{fig:speedEuLa} depicts the average pedestrian velocity field for the considered queries.  The walking speed varies in space, and its contours are roughly transversal with respect to the walking direction.  In both Lagrangian and Eulerian points of view, pedestrians going 2R walk slower than the pedestrians going 2L.  In Lagrangian queries the speed field is not symmetric to the middle line of the corridor.  Pedestrians walk in a higher speed at the later part of their walk in the corridor, but before they arrive the next flight of stairs they slow down again.  A speed drop of about  $30\%$ is measured in our observation window.  In both 2L and 2R cases the deceleration phase when pedestrians approach  the next flight of stairs is shorter than the acceleration phase when they arrive the landing.  In the Eulerian perspective pedestrian speed is lower than what is found using Lagrangian queries.  Since a single pedestrian in the Eulerian query may have co-flow or counter-flow encounters during their walk in the corridor, his/her speed may reduce to reflect such a situation. Also in the Eulerian point of view the asymmetry of acceleration and deceleration when entering/leaving the landing is greatly reduced, although still visible. 

\begin{figure}[ht!]   
\begin{center}
\subfigure[]{\includegraphics[width=.4715\textwidth,trim=0 .2cm 2.8cm 0, clip=true]{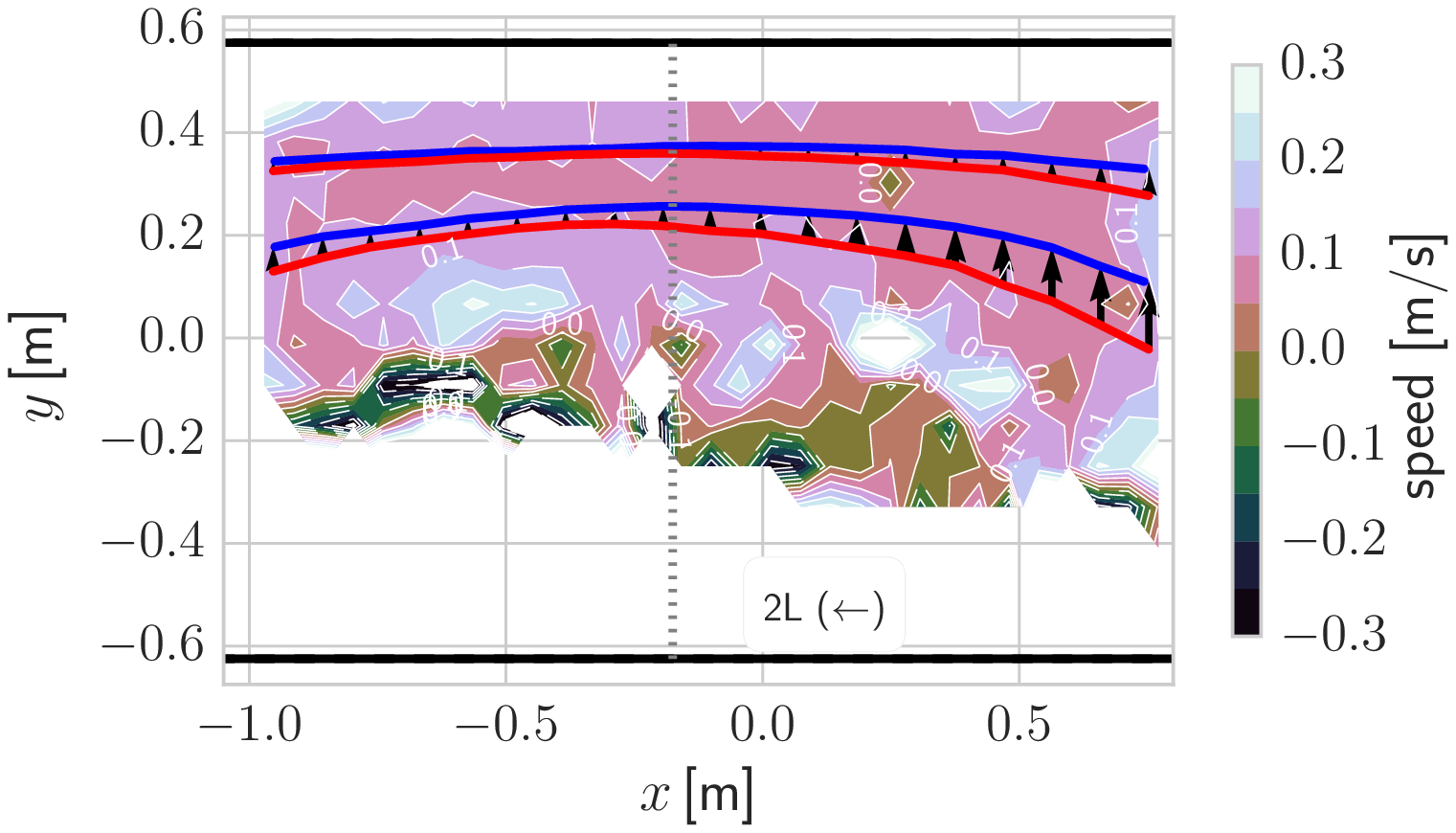}}
\subfigure[]{\includegraphics[width=.49\textwidth, trim=2.3cm .2cm 0cm 0, clip=true]{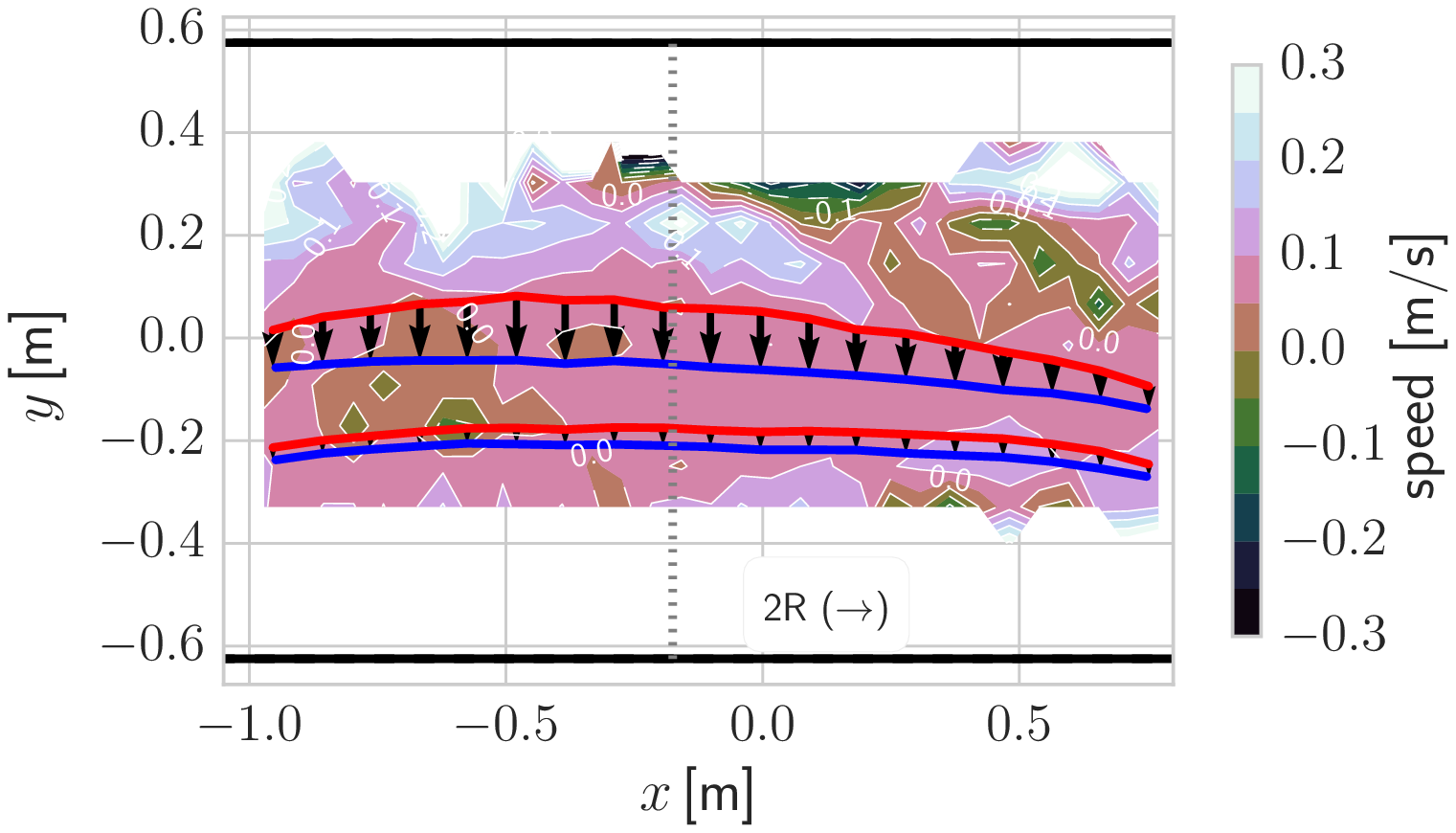}}
\end{center}
\caption{Differences in the preferred position bands and in the average velocity fields of two pedestrians in the counter-flow configuration from Lagrangian and Eulerian queries.  The red lines represent the preferred position bands from the Lagrangian queries, the blue lines represent that from the Eulerian queries, and black arrows indicate the change of the results from the Lagrangian queries to Eulerian queries.  The difference of preferred position band is much larger in this counter-flowing configuration than in the single pedestrian case. In the Eulerian case preferred position bands are much more to the relative right side of the corridor due to the potential higher traffic. The underlying maps presents the difference in average speed field calculated by subtracting average speed field of Eulerian queries from that of Lagrangian queries.  In most of the regions the average speed from Lagrangian queries is larger due to that fact that in Eulerian queries the pedestrians are already sharing the corridor at the moment of recording and the avoidance mechanism has taken effect.}
\label{fig:speedLa-Eu-two-ped} 
\end{figure}

In the same spirit, we compute the preferred position bands and the average velocity field for the case of two pedestrians walking in the counter-flow condition.  \fref{fig:speedLa-Eu-two-ped} shows the difference between the Lagrangian and Eulerian queries.   It is clear that the difference in the preferred position bands from the two perspective is even larger compared with the single pedestrian case (cf. \fref{fig:singlePedEuLa}).  The Eulerian queries of this condition gives frames of  exactly two pedestrians who walk in opposite directions.  Hence the pedestrians have already seen each other, and the avoidance mechanism (topic of \sref{sect:lagr}) is in effect.  Also these pedestrians may encounter more people during their walk in the corridor.  Compared with the Lagrangian queries of this condition where the pedestrians may not have met their counter part yet or the other person has left, the Eulerian queried counter-flow trajectories show  a greater avoidance effect.  We can do the same two set of queries for the average velocity fields and calculate the difference, as shown in the underlying map in \fref{fig:speedLa-Eu-two-ped}.  In most of the region the average velocity field of the Lagrangian queries is in a larger magnitude (so the difference is positive).  The aforementioned reasons are likely to contribute to this difference as well.

\subsection{Lagrangian analysis of pair-wise interactions} \label{sect:lagr}
The last step of our analysis considers  Lagrangian scenarios involving undisturbed pedestrians and counter-flowing pairs (i.e. the cases P1, P2, P3-P4 in \fref{fig:graph-lagr-constr}). The asymmetries, here discussed quantitatively, are not limited to positions and velocities and include, in a social-force modeling~\cite{helbing1995PRE} perspective, exchanged ``interaction forces'' (here rather accelerations).

\begin{figure}[ht!]
\begin{center}
\subfigure[]{\includegraphics[width=.8\textwidth, clip=true]{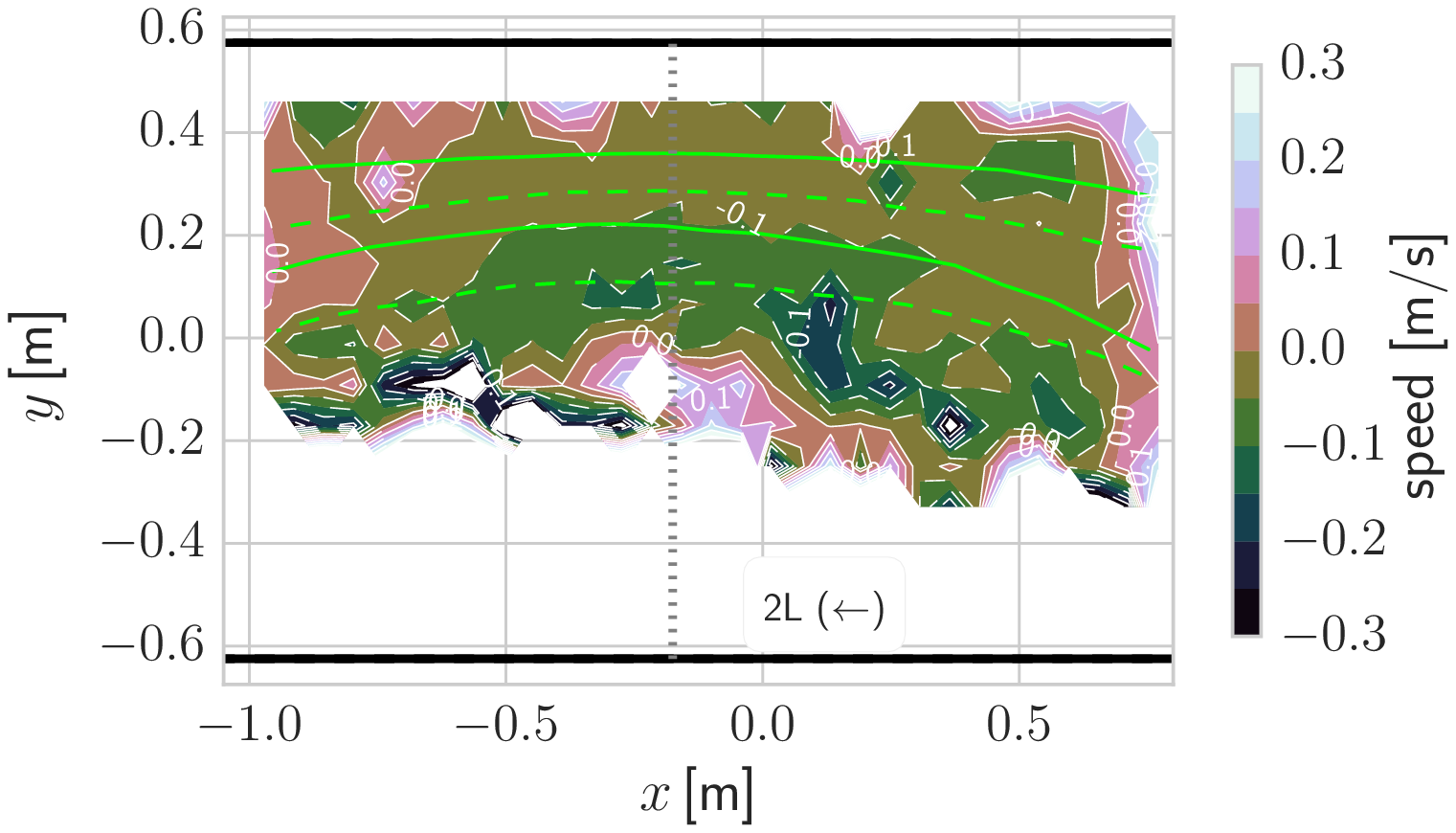}}
\subfigure[]{\includegraphics[width=.8\textwidth, clip=true]{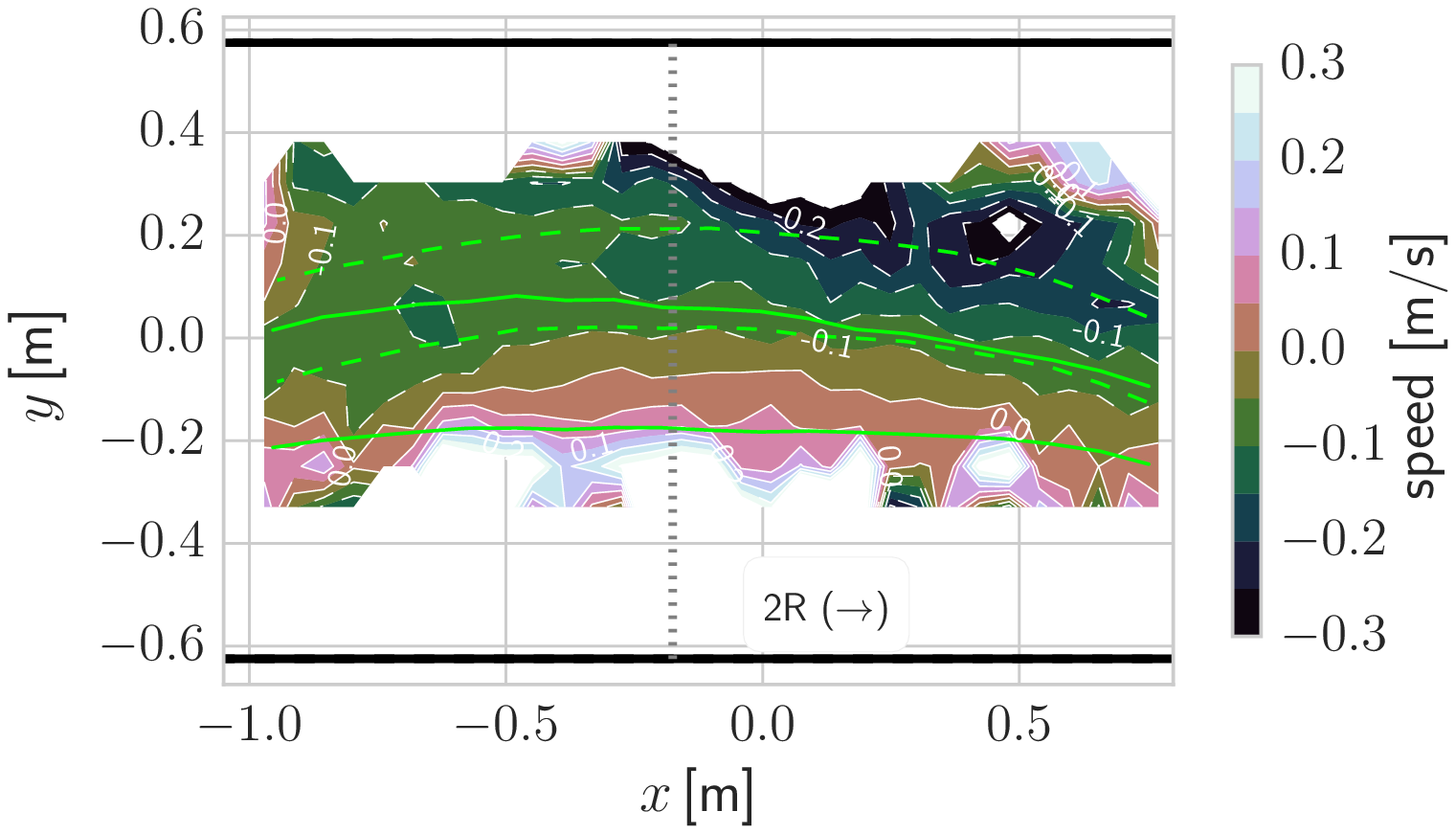}}
\end{center}
\caption{Preferred position bands and difference of average speed between counter-flow pairs  and undisturbed pedestrians from Lagrangian queries.  The dashed green lines represent the preferred position bands when pedestrians walking undisturbed, and the solid green lines represent the preferred position bands for counter-flow pairs.  The preferred position bands are shifted toward the relative right side of the corridor in the counter-flowing configuration due to collision avoidance.  The underlying map shows the average velocity fields of counter-flow pairs subtracts the average speed fields of undisturbed pedestrians. (a) case 2L, (b) case 2R.  In most of the space the difference is negative since counter-flowing pedestrians slow down during the encounter with another pedestrian. }
\label{fig:2counterflow}  
\end{figure}

\begin{figure}[ht!]
\begin{center}
\subfigure[]{\includegraphics[width=.3\textwidth]{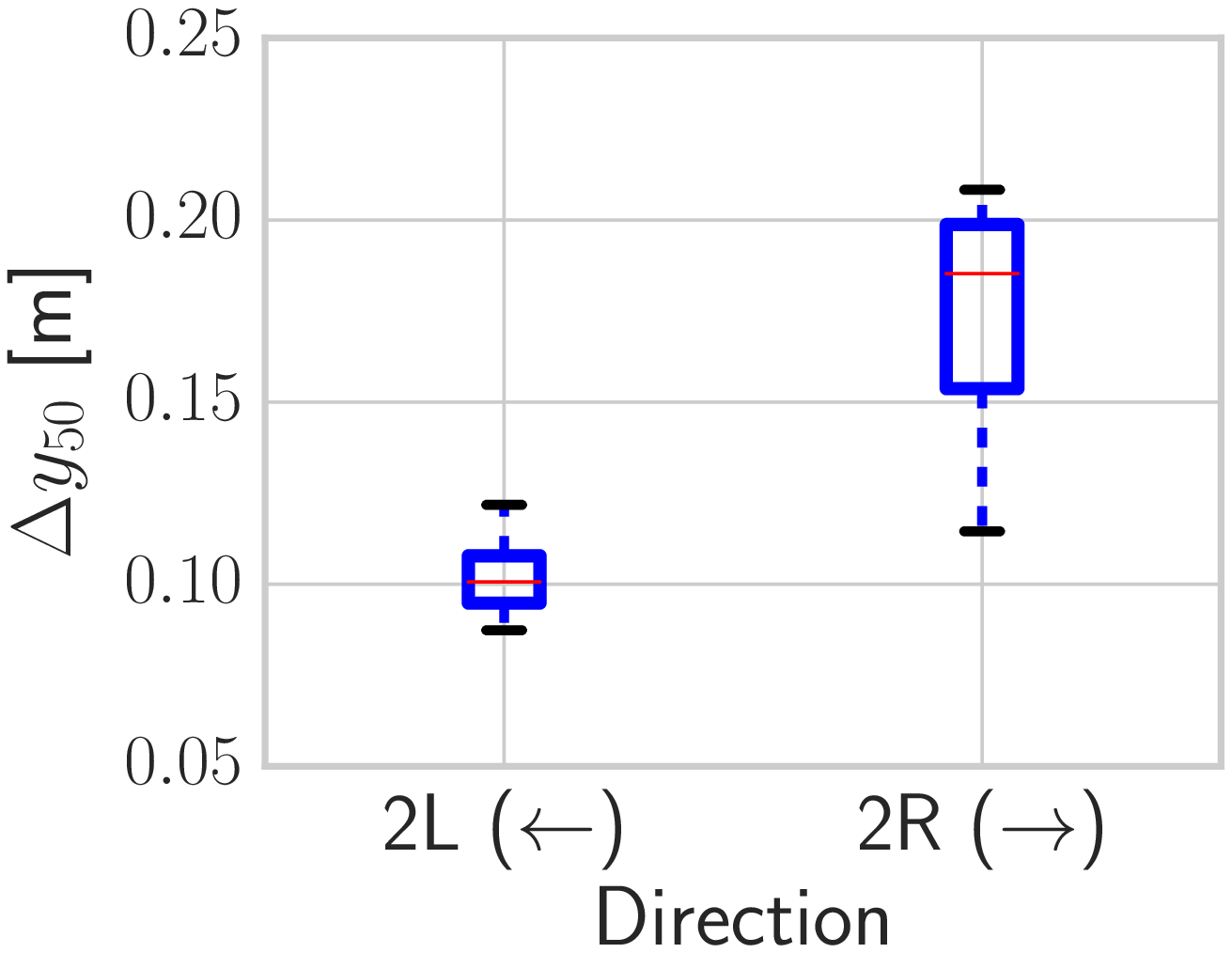}} 
\subfigure[]{\includegraphics[width=.3\textwidth]{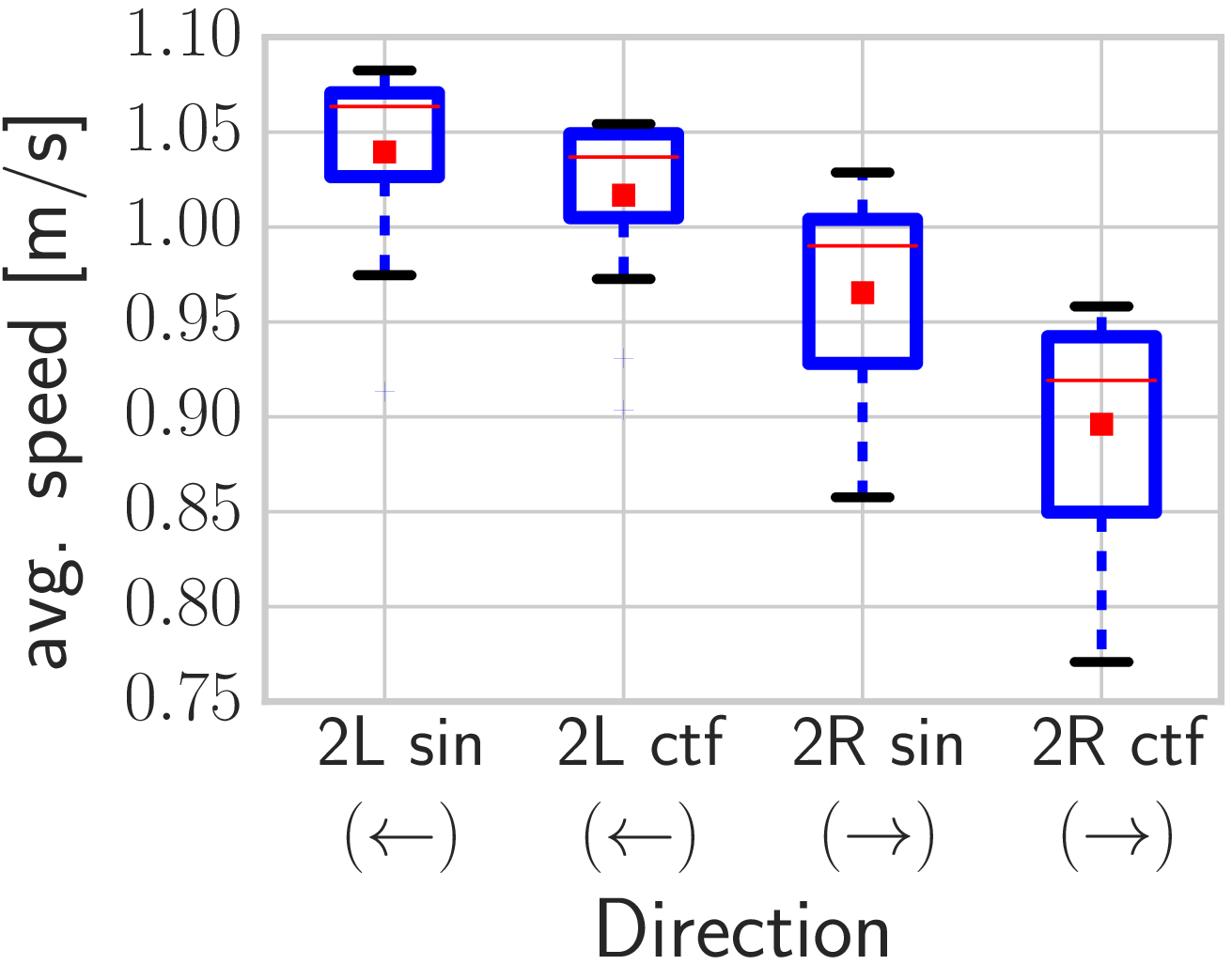}}  
\subfigure[]{\includegraphics[width=.3\textwidth]{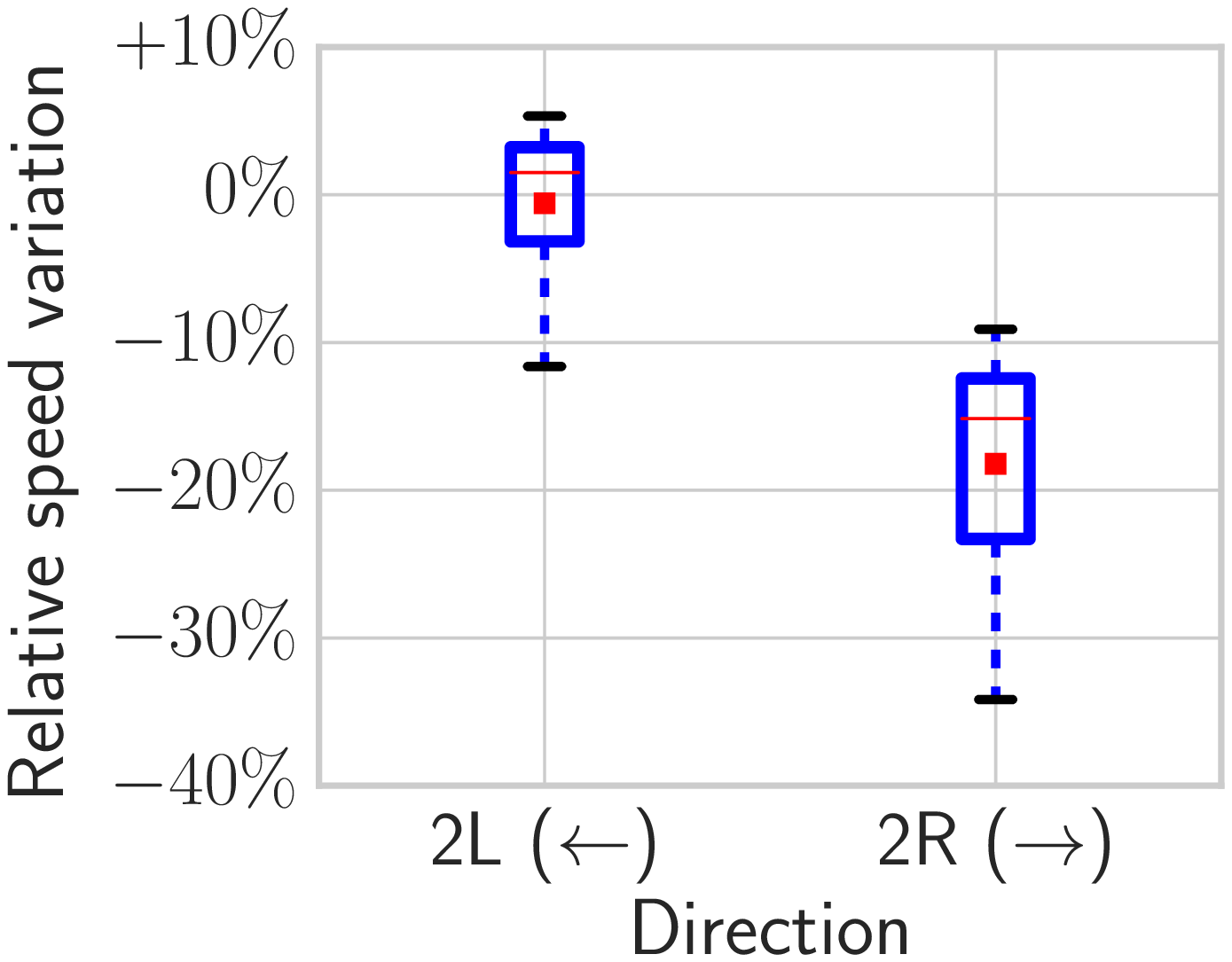}} 
\end{center}
\caption{Quantitative comparisons between the undisturbed dynamics and the counter-flow of two individuals in terms of position and velocity differences. (a) Displacement in absolute value of the layers of preferred positions from the undisturbed pedestrian case to the two pedestrians in counter-flow (cf. \fref{fig:2counterflow}). We evaluate the transversal displacement of the median line of the path $y_{x,50}$ (cf. explanation in  \aref{app:tech}). We consider  $\Delta y_{x,50}^{2L} = (y_{x,50}^{counterflow,2L} - y_{x,50}^{single,2L})$ in the 2L case and $\Delta y_{x,50}^{2R} = -(y_{x,50}^{counterflow,2R} - y_{x,50}^{single,2R})$ in the 2R case. The box-plots report the distributions of $\Delta y_{x,50}^{2L}$ and of $\Delta y_{x,50}^{2R}$ across the landing span.  The 2R pedestrians moved much more to the relative right when sharing the corridor with another pedestrian coming their way.  (b) Comparison of walking speeds for undisturbed pedestrians and pair pedestrians in counter flow.  We evaluate the average speed at each span-wise location $x$, and for every direction we consider the relative  difference of such velocity between undisturbed case and counter-flow. The distribution of the relative difference is reported by the box-plots. Descending (2L) pedestrians walk faster than 2R pedestrian in both undisturbed and counter-flowing situations.  Although in both directions pedestrians slow down when in counter-flow, the 2R pedestrians slowed down much more, which can also be seen in panel (c).  (c) Distribution of relative speed variation comparing undisturbed pedestrians and counter-flows of two. (a,b,c) The box-plots limit the first and third quartiles of the distributions, the whiskers identify the 5$^{th}$ and 95$^{th}$ percentiles. Red line reports the median and red dots the average values.
} 
\label{fig:boxplots}
\end{figure}

 Direction-dependent differences, considered for undisturbed pedestrians in \fref{fig:singlePedEuLa} and \fref{fig:singlePedEuLaPath-decomp}{cd}, increase when the presence of one other pedestrian with opposite direction triggers the avoidance mechanism (i.e., in a counter-flow. Case  P3 -- P4 in \fref{fig:graph-lagr-constr}). In this condition, the  paths are shifted to the relative right to avoid collision. Contrary to the single pedestrian case these preferred position bands have no overlap (cf. \fref{fig:2counterflow} and \fref{fig:singlePedEuLa}), furthermore they are not symmetric with respect to the vertical axis in the landing center ($x\approx -0.1\,$m). In both 2L and 2R cases, bands are wider near the entrance side with similar distribution to  the undisturbed pedestrian case. Moving across the landing, the bands constrict and shift toward the relative right-hand side, possible due to encounters of other pedestrians.  The displacement of the preferred position bands features direction-related asymmetries: on average the rigid translation of the band is \textit{ca.} $40\%$ larger in the 2R case than in 2L case.  For pedestrians going to the left the preferred position band shifts almost rigidly to the relative right showing a displacement of \textit{ca.} $10\,$cm. For pedestrians going to the right, instead, the preferred position band has a deformation. The band axis shifts of \textit{ca.} $18\,$cm. (cf. \fref{fig:2counterflow} and \fref{fig:boxplots}{a}).

We observe a  drop in the walking speed in comparison with the undisturbed pedestrians, especially around the central horizontal axis ($y\approx 0\,$m) where collisions may potentially occur. Higher walking speed are reached at the  relative right hand side of the pedestrians,  where collisions are mostly avoided (cf. the negative difference in the average speed fields in \fref{fig:2counterflow}).   \fref{fig:boxplots}{b}{c} show quantitative measurements of the speed reduction.  In both undisturbed and counter-flow pedestrian cases, pedestrians walking to the right walk slower than the pedestrians walking to the left.  Furthermore, when encountering another pedestrian coming from the opposite direction, the 2R pedestrians slow down $18\%$ in average, while the descending pedestrians (2L) has a almost negligible reduction.

\begin{figure}[ht!]
\begin{center}
\includegraphics[width=.9\textwidth]{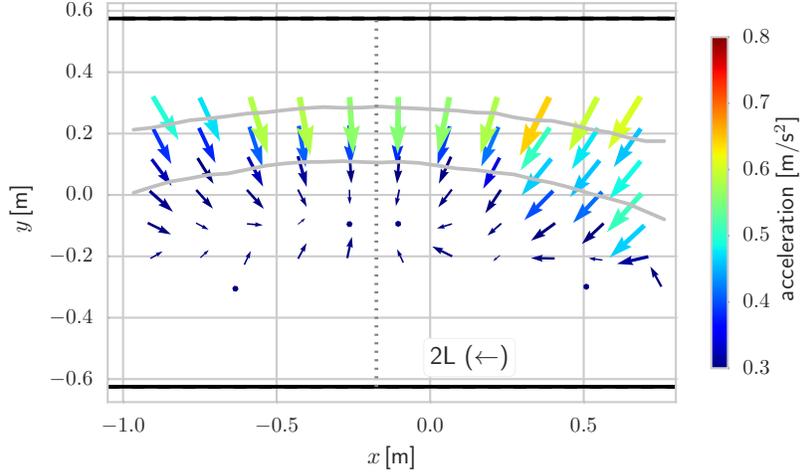}
\end{center}
\caption{Average acceleration field for undisturbed pedestrians walking from right to left. The curved pedestrian motion following the U-shape of the corridor determines a centripetal acceleration. We measure an almost central acceleration field pointing to \textit{ca.} $(x=-0.25,y=-0.10)\,$m. The acceleration field for pedestrians going to the right is analogous, thus not repeated.}
\label{fig:displacement-distribution} 
\end{figure}   
     
\begin{figure}[ht!]        
\begin{center}
\subfigure[]{\includegraphics[width=.9\textwidth]{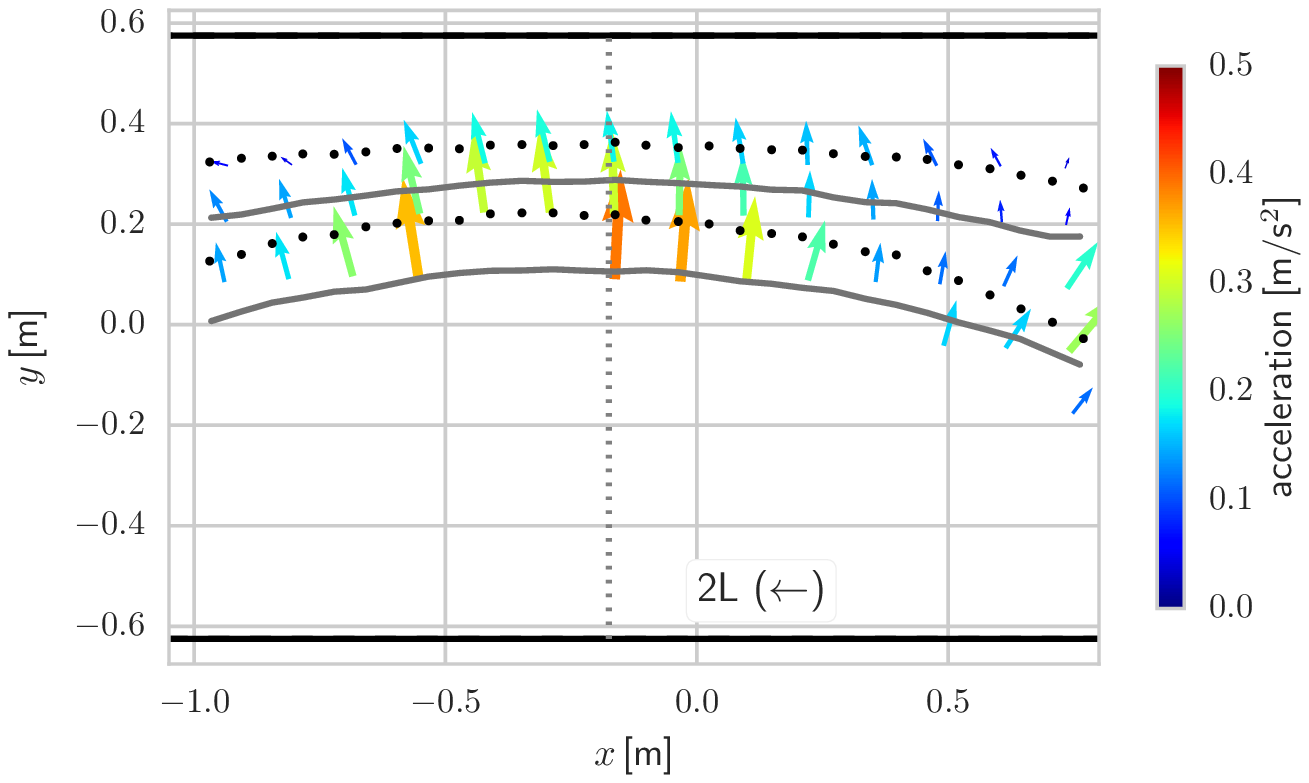}}
\subfigure[]{\includegraphics[width=.9\textwidth]{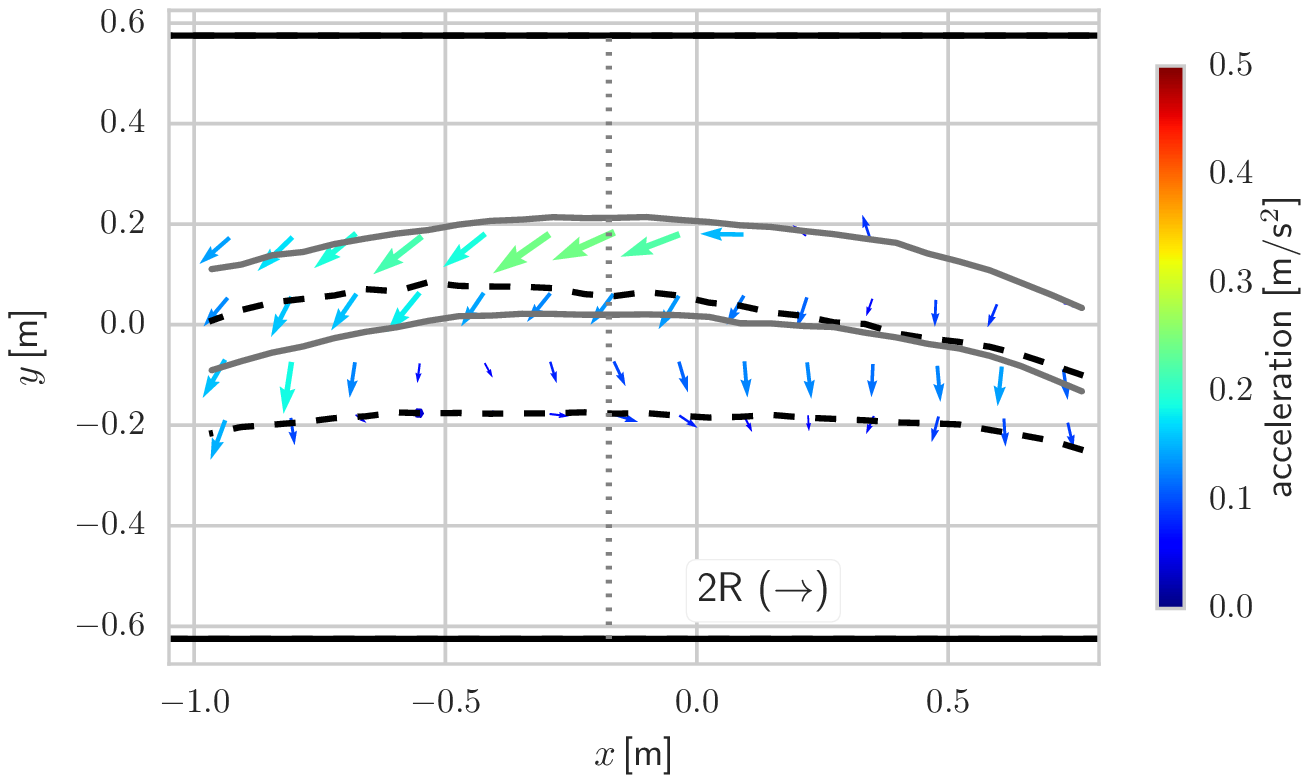}}   
\end{center}  
\caption{Average avoidance acceleration field (cf. \eref{eq:interaction-acceleration}) for two pedestrians having opposite velocities (counter-flowing). (a) acceleration field for pedestrians going from right to left; (b) acceleration field for pedestrians going from left to right. In both cases the fields yield sidesteps on the relative right and a longitudinal speed reduction in the entering half of the landing. In the preferred walking band of undisturbed pedestrians (solid line) deceleration are stronger. The dashed line reports the preferred walking band for the two pedestrians case. 
  }  
\label{fig:interaction-force}
\end{figure}

Acceleration fields help the interpretation of the position and average velocity fields in \fref{fig:2counterflow}. As pedestrians follow curved trajectories, they experience a centripetal acceleration even when moving undisturbed (cf. \fref{fig:displacement-distribution} for the average acceleration field of pedestrians walking from the right to the left). Lagrangian queries are here paramount as to quantify fields without perturbations from pedestrians just met or to be met in the landing. Let $\vec{a}_{p,1}$ be the acceleration of undisturbed pedestrians. To estimate the average acceleration field in avoidance, we follow a social force-like~\cite{helbing1995PRE} approach decomposing the acceleration of a pedestrian $\vec{a}_p$ as  
\begin{equation}
\vec{a}_p = \vec{a}_{p,d} + \vec{a}_{p,i},
\end{equation}
where $\vec{a}_{p,d}$ denotes the \textit{desired} component of the acceleration, thus independent on other pedestrians, and $\vec{a}_{p,i}$ is the perturbation to $\vec{a}_{p,d}$ due to the avoidance \textit{interaction}. In social force models, $\vec{a}_{p,d}$ is typically a relaxation force toward a given (desired) velocity field. It is reasonable to assume that, at least  on average,  the force $\vec{a}_{p,d}$ can be approximated as $\vec{a}_{p,1}$. Thus, we extract the average interaction acceleration as   
\begin{equation}\label{eq:interaction-acceleration}
\langle\vec{a}_{p,i}\rangle = \langle\vec{a}_p \rangle -  \langle\vec{a}_{p,d}\rangle \approx \langle\vec{a}_p \rangle -  \langle\vec{a}_{p,1}\rangle,
\end{equation}
where $\langle \vec{a} \rangle$ is a local spatial average of $\vec{a}$ (cf. \aref{app:tech} for technical details). We report the avoidance acceleration fields for pedestrians going to the left and to the right in \fref{fig:interaction-force}.

In both 2L and 2R cases the accelerations point toward the relative right following the drift of trajectories thus the displacement of the preferred position bands. Strong longitudinal decelerations for collision avoidance are visible in the 2R case close by the entering side (left end).  However it is noticed that pedestrians going to the left mostly avoid collision by moving to the relative right without changing the forward speed (longitudinal velocity).

\section{Discussion}\label{sect:concl}
Pedestrian dynamics measurements acquired in real-life conditions are largely heterogeneous due to the natural variability of flow conditions. This makes data selection paramount, as accidental aggregations of data from heterogeneous flows may yield biased statistical measurements leading to improper conclusions. This demands for methodologies to query homogeneous datasets from measurements. 

In this paper we cross compared pedestrian dynamics data from a large experimental dataset, that we collected via a year-long measurement campaign on a staircase landing. From the physical point of view we commented on the asymmetries of the pedestrian dynamics depending on flow conditions, here identified with the number of pedestrians and their walking directions. The U-shape of the landing combines with functional differences of walking directions (pedestrians walking to the left are going to a lower level in the building, the opposite happens for pedestrians going to the right)  yielding asymmetries at the levels of preferred positions, velocities and also avoidance accelerations (social interaction forces). The quantitative differences found include, beside higher velocities for pedestrians descending, larger influences to the walking patterns of ascending pedestrians in presence of  pedestrians walking in counter-flow. We observed a strong walking side preference towards the driving side, indeed  an influence of the landing shape on this cannot be excluded. In these conditions ascending pedestrians, typically walking in the inner side, may have a different sight range on the environment than individual descending. This aspect can play also a role in the asymmetries measured.

The previous comparisons stimulated methodological investigations too. We recognised the difference between querying our dataset for homogeneous flow conditions at the frame level (Eulerian querying) or at the trajectory level (Lagrangian querying). Although querying at the frame level is more immediate when dealing with continuous measurements, pedestrian interactions affect pedestrian trajectories thoroughly. While this might not be true on large length scales, it certainly holds for our recording window. A pedestrian observed alone remains affected by previous passes-by of other individuals (already disappeared from the observation window) or, since sight likely extends beyond our observation window, avoidance maneuvers may already play a role before a second pedestrian enters in the scene. In this respect, the following observations hold:
\begin{itemize}
\item the graph connections are built after simultaneous appearance of pedestrians in our observation window. In a sense, we assume that interactions among pedestrians are limited to connected components. In principle, interactions might happen outside our recording windows and still play a role in the observed dynamics. Although this aspect is hard to assess, we expect that because of the geometry of the landing it played negligible effects. In other words we assume that if two pedestrian interact then, at a point, both appeared in our observation window.
 \item The spatial scale of our geometry is certainly relevant. Although it is reasonable to assume that two pedestrians appearing together in our observation window play a reciprocal influence on their dynamics, the same would not hold for larger geometrical settings (e.g. spanning beyond typical interaction ranges). Generalizing the graph structure including geometric distances might help in treating such cases.
\end{itemize}

\begin{appendix}

\section{Preferred walking layers, speed and acceleration fields}\label{app:tech}
We evaluate preferred walking layers, speed and acceleration fields via a spatial binning of the measurements from homogeneous sets of  trajectories (cf. \sref{sect:query}). Given a homogeneous set of trajectories every detection $d$ has form
\begin{equation}
d = (t,p,x,y,u,w,a_x,a_y)
\end{equation}
where $t$ is the detection time, $p$ is an unique index for the detected pedestrian, $(x,y)$ is the position of the pedestrian at time $t$, $\vec{v}=(u,w)$ his/her velocity, $\vec{a} = (a_x,a_y)$ his/her acceleration. 

To evaluate the preferred walking layer we extend the approach suggested in~\cite{corbetta2014TRP}. We bin the detection set $\{d\}$ with respect to the longitudinal position $x$ between $x = -1\,$m and $x=0.8\,$m in $40$ equal bins. For each bin we consider the distribution of transversal positions $y_x$ (where the $x$ subscript indicates the dependence on the bin), and we take the $15^{th}$ and $85^{th}$ percentiles (indicated by $y_{x,15}$ and $y_{x,85}$) of the distribution to define the preferred position band.

We evaluate the displacement of the preferred position bands in \fref{fig:boxplots}{a} by considering in each bin the $50^{th}$ percentiles of $y_x$: $y_{x,50}$. We compute the difference between the value for   pedestrians undisturbed ($y_{x,50}^{single,2L}$, for the 2L case) and walking with one other individual in opposite direction ($y_{x,50}^{counterflow,2L}$). We consider the distribution of the quantity obtained  ($\Delta y_{x,50}^{2L} = (y_{x,50}^{counterflow,2L} - y_{x,50}^{single,2L})$).

Velocity and acceleration fields are defined after a binning with respect to $x$ and $y$. For each bin we take respectively the average speed $\langle\sqrt{u^2+v^2}\rangle$ and the average acceleration $\langle\vec{a}\rangle = (\langle a_x\rangle,\langle a_y\rangle)$. For the velocity fields we employed $32$ bins  in $x$ direction within $[-0.8,1.0]\,$m and $20$ bins in $y$ direction within $[-1.0,0.5]\,$m. For the acceleration fields we employed $20$ bins  in $x$ direction within $[-0.8,1.0]\,$m and $20$ bins in $y$ direction within $[-0.4,0.4]\,$m. 

\end{appendix}

\begin{acknowledgement}
We  thank A. Holten and G. Oerlemans (Eindhoven, NL) for their help in the establishment of the measurement setup at Eindhoven University of Technology and  A. Liberzon (Tel Aviv, IL) for his help in the adaptation of the OpenPTV library. We acknowledge the support from the Brilliant Streets research program of the Intelligent Lighting Institute at the Eindhoven University of Technology, NL. This work is part of the JSTP research programme ``Vision driven visitor behaviour analysis and crowd management" with project number 341-10-001, which is financed by the Netherlands Organisation for Scientific Research (NWO).  Support from COST Action MP1305 ``Flowing Matter'' is also kindly acknowledged.
\end{acknowledgement}

\bibliographystyle{cdbibstyle} \bibliography{bibliog}

\end{document}